\documentclass[12pt]{article}

\usepackage{graphicx, amsmath, amsthm, amsfonts, bigints}
\usepackage{color,ulem, psfrag}
\usepackage{comment}
\usepackage{url}
\usepackage[ruled]{algorithm2e}
\usepackage{xcolor}

\newcommand {\dn}[1] {\boldsymbol #1}
\newcommand {\mt}[1] {\mathrm #1}

\newcommand {\be} {\begin {equation} }
\newcommand {\ee} {\end {equation} }
\newcommand {\bmath} {\begin {displaymath} }
\newcommand {\emath} {\end {displaymath} }

\newtheoremstyle{mytheoremstyle}
{3pt}
{3pt}
{\itshape}
{}
{\scshape}
{:}
{.5em}
{}
\theoremstyle{mytheoremstyle}

\usepackage{harvard} 
\newcommand{\citep}{\citeasnoun}  
\newcommand{\citea}{\citeaffixed}    

\begin{document}
\normalem
\renewcommand{\baselinestretch}{1.55}\normalsize

\title{\vspace*{-0.5in}
\textbf{Power-Expected-Posterior Priors as Mixtures of $g$-Priors in Normal Linear Models}}

\author{
D.~Fouskakis\thanks{D.~Fouskakis is with the Department of
Mathematics, National Technical University of Athens, Greece; email
\texttt{fouskakis@math.ntua.gr}} \
and  I.~Ntzoufras\thanks{I.~Ntzoufras is with the Department of
Statistics, Athens University of Economics and Business, Greece; email
\texttt{ntzoufras@aueb.gr}} 
}

\date{}

\maketitle

\noindent
\textbf{Summary:} One of the main approaches used to construct prior distributions for objective Bayes
methods is the concept of random imaginary observations.  Under this setup, the expected-posterior prior (EPP) offers several advantages, among which it has a nice and simple interpretation and provides an effective way to establish compatibility of priors among models. In this paper, we study the power-expected-posterior prior as a generalization to the EPP in objective Bayesian model selection under normal linear models. We prove that it can be represented as a mixture of $g$-prior, like a wide range of prior distributions under normal linear models, and thus posterior distributions and Bayes factors are derived in closed form, keeping therefore its computational tractability. Following this result, we can naturally prove that desiderata (criteria for objective Bayesian model comparison) hold for the PEP prior. 
Comparisons with other mixtures of $g$-prior are made and results are presented in simulated and real-life datasets.    
\vspace*{0.15in}


\section{Introduction}

Let $\dn{y}=(y_1, \dots, y_n)^T$ denote some available observations. Under the objective Bayesian perspective, suppose we wish to compare the following two models (or hypotheses):
 \begin{eqnarray}
 \label{hc}
 \mbox {model } M_0: f(\dn{y} | \dn{\theta}_0, M_0),~~ \dn{\theta}_0 \in \Theta_0 \nonumber \\
 \mbox {model } M_1: f(\dn{y} | \dn{\theta}_1, M_1), ~~ \dn{\theta}_1 \in \Theta_1, 
 \end{eqnarray}
 where $\dn{\theta}_0$ and $\dn{\theta}_1$ are unknown, model specific, parameters. Let further suppose that $M_0$ is nested in $M_1$. By $\pi_{\ell}^N(\dn{\theta}_\ell)$, for $\ell \in \{0,1\}$, we denote the baseline prior of $\dn{\theta}_\ell$ under model $M_\ell$. Here, as a baseline prior we consider any prior that will express low information, for example the reference prior; see \citep{berger_bernardo_sun_2009}. These reference priors are typically improper, resulting in a Bayes factor when comparing $M_0$ to $M_1$ which typically cannot be determined due to the unknown normalizing constants of these improper priors. 

In order to specify these unknown normalizing constants, 
\citep{perez_berger_2002} developed priors 
through utilization of the device of ``imaginary training samples". If we denote by $\dn{ y }^*$ the imaginary training sample, of size $n^*$, they defined the
expected-posterior prior (EPP) for the parameter vector $\dn{ \theta }_\ell$, of model $M_\ell$, as
\begin{equation} 
\label{epp} \pi^{EPP}_\ell ( \dn{ \theta }_\ell ) = \int
\pi_\ell^N ( \dn{ \theta }_\ell | \dn{ y }^* ) \, m^* ( \dn{ y }^* ) \, d
\dn{ y }^* \, , 
\end{equation} 
where $\pi_\ell^N ( \dn{ \theta }_\ell | \dn{y }^* )$ is the posterior of $\dn{ \theta }_\ell$ for model $M_\ell$ using the baseline prior $\pi_\ell^N ( \dn{ \theta }_\ell )$ and data $\dn{ y }^*$ and 
$m^* ( \dn{ y }^* )$ is a reference predictive distribution generating sets of imaginary data. 
A usual  choice of $m^*$ is $m^* ( \dn{ y }^* ) = m_0^N ( \dn{ y }^* )
\equiv f ( \dn{ y }^* | M_0)$, i.e. the marginal likelihood,
evaluated at $\dn{ y }^*$, for the simplest model $M_0$ under the baseline
prior $\pi_0^N ( \dn{ \theta }_0 )$.  Then model $M_0$ is called the reference model. EPP offers several advantages, among which it has a nice interpretation and also provides an effective way to establish compatibility of priors among models \cite{consonni_veronese_2008}. Furthermore this selection makes the EPP approach essentially equivalent to the arithmetic intrinsic Bayes factor approach of \citep{berger_pericchi_96b} since it is straightforward to prove that the EPP can be written equivalently as
\begin{equation*} 
\pi_\ell^{EPP} ( \dn{\theta}_\ell ) = \pi^N_\ell( \dn{\theta}_\ell) E_{Y^*}^{M_\ell} \left[ \frac{m^N_0( \dn{y}^* )}{m^N_\ell( \dn{y}^* )}\right]
= \pi^N_\ell( \dn{\theta}_\ell) \int \frac{m^N_0( \dn{y}^* )}{m^N_\ell( \dn{y}^* )} f( \dn{y}^* | \dn{\theta}_\ell, M_\ell ) d\dn{y}^*.
\label{epp_in}
\end{equation*}

When information on covariates is also available, under the EPP methodology, imaginary design matrices $\mt{ X }^*$ with $n^*$ rows should also be introduced. The selection of a \textit{minimal}
training sample size $n^*$ has been proposed by \citep{berger_pericchi_2004}, to make the information
content of the prior as small as possible, and this is an appealing idea. Then $\mt{ X }^*$ can be extracted from the original design matrix $\mt{ X }$, by randomly selecting $n^*$ from the $n$ rows. 
  
To diminish the effect of training samples, \citep{fouskakis_et.al_2013} generalized the EPP approach, by 
introducing the \textit{power-expected-posterior} (PEP) priors, combining
ideas from the power-prior approach of \citep{ibrahim_chen_2000} and the
unit-information-prior approach of \citep{kass_wasserman_95}. 
As a first step, the likelihoods involved in the EPP formula are raised to the
power $1/\delta$ $(\delta \geq 1)$ and then are density-normalized; for a discussion on the density--normalization and different versions of the PEP priors when this normalization leads to no standard forms see \citep{fouskakis_etal_2017}. This power
parameter $\delta$ could be then set equal to the size of the training sample $n^*$, to represent information equal to one
data point. In \citep{fouskakis_et.al_2013} the authors further set $n^* = n$; this choice gives rise 
to significant advantages, for example when covariates are available it results in the automatic choice $\mt{ X }^* = \mt{ X }$ and therefore the selection of a training sample and its effects on the
posterior model comparison is avoided, while still holding the prior information
content at one data point. 

Specifically, for the model selection problem (\ref{hc}), the PEP prior is defined as 
\be 
\pi_\ell^{PEP} ( \dn{\theta}_\ell | \delta ) \equiv  \pi_\ell^{PEP} ( \dn{\theta}_\ell ) = 
\int \pi_\ell^N( \dn{\theta}_\ell | \dn{y}^*, \delta) m^*( \dn{y}^*| \delta) d \dn{y}^*,
\label{pep}
\ee
with 
\begin{eqnarray}
\pi_\ell^N( \dn{\theta}_\ell | \dn{y}^*, \delta) &\propto& f( \dn{y}^* | \dn{\theta}_\ell, \delta, M_\ell ) \pi_\ell^N ( \dn{\theta}_\ell  ) \label{pep1}\\ 
f( \dn{y}^* | \dn{\theta}_\ell, \delta, M_\ell ) &=& \frac{ f( \dn{y}^* | \dn{\theta}_\ell,  M_\ell )^{1/\delta} } 
{ \int f( \dn{y}^* | \dn{\theta}_\ell,  M_\ell )^{1/\delta} d\dn{y}^* }~.\label{pep2}
\end{eqnarray}
As before we choose 
\be
m^*( \dn{y}^*| \delta) = m^N_0( \dn{y}^* | \delta ) = 
\int f( \dn{y}^* | \dn{\theta}_0, \delta, M_0 ) \pi^N_0( \dn{\theta}_0 ) d \dn{\theta}_0 ~, \label{pep3}
\ee 
where $f( \dn{y}^* | \dn{\theta}_0, \delta, M_0 )$ is given by \eqref{pep2} for $\ell = 0$ (i.e. the null/reference model).

In this work we show, using sufficient statistics (see \citep{fouskakis_2019} for the equivalent definitions of EPP and PEP prior using sufficient statistics), that the PEP prior (and therefore also the EPP), in normal linear model comparison, can be expressed as a mixture of $g$-priors, centred around null models. 
This has the advantage that posterior distributions, as well as, marginal likelihoods are available in closed form and desiderata (criteria for objective Bayesian model comparison), see \citep{Bayarri_etal_2012}, hold.
We compare the PEP prior with other scale normal mixture priors, we present prior summaries for model parameters and we derive posterior distributions as well as marginal likelihoods. Emphasis is given to the Bayesian inference of the shrinkage parameter which is also involved in Bayesian model averaging estimation. Finally we present results from a simulation study, as well as from a real--life dataset. 


\section{PEP Prior as a Mixture of Normal Distribution in Normal Linear Model Comparison} 

Let $\dn{y}=(y_1, \dots, y_n)^T$ be a random sample. We would like to compare the nested models:
\begin{eqnarray*}
&&H_0 : \mbox {model } M_0: \mbox{Normal}(\dn{y} | \mt{X}_0\dn{\beta}_0, \sigma_0^2), ~~ \pi_0^N(\dn{\beta}_0, \sigma_0) =c_0 \pi_0^U(\dn{\beta}_0, \sigma_0) = c_0 \sigma_0^{-{(1+d_0)}}\\
&vs.&H_1 : \mbox {model } M_1: \mbox{Normal}(\dn{y} | \mt{X}_1\dn{\beta}_1, \sigma_1^2), ~~ \pi_1^N(\dn{\beta}_1, \sigma_1) = c_1 \pi_1^U(\dn{\beta}_1, \sigma_1) = c_1 \sigma_1{^{-(1+d_1)}}
\end{eqnarray*}
where $c_0$ and $c_1$ are the unknown normalizing constants of $\pi_0^U(\dn{\beta}_0, \sigma_0)$ and $\pi_1^U(\dn{\beta}_1, \sigma_1)$ respectively, $\mt{X}_0$ is an $(n \times k_0)$ design matrix under model $M_0$, $\mt{X}_1$ is an $(n \times k_1)$ design matrix under model $M_1$, $k_0<k_1<n$ and $M_0$ is nested in $M_1$.  Furthermore let $\dn{\beta}_1 = \left(\dn{\beta}_0^T, \dn{\beta}_{e_1}^T \right)^T$, $\mt{X}_1 = \left[\mt{X_0} | \mt{X}_{e_1} \right]$, $\mt{P}_0 = \mt{X}_0\left({\mt{X}_0}^T \mt{X}_0 \right)^{-1} {\mt{X}_0}^T$ and $\mt{P}_1 = \mt{X}_1\left({\mt{X}_1}^T \mt{X}_1 \right)^{-1} {\mt{X}_1}^T$. All matrices are assumed to be of full rank. Usual choices for $d_0$ and $d_1$ are $d_0=d_1=0$ (resulting to the reference prior) or $d_0=k_0$ and $d_1=k_1$ (resulting to the dependence Jeffreys prior).

In the above comparison we assume that model $M_0$ is nested in model $M_1$, so that $k_0<k_1$ and thus we henceforth assume that $\dn{\beta}_1 = \left(\dn{\beta}_0^T, \dn{\beta}_{e_1}^T \right)^T$, so that $\dn{\beta}_0$ is a parameter ``common" between the two models, where $\dn{\beta}_{e_1}$ is model specific. 
The use of a ``common" parameter $\dn{\beta}_0$ in nested model comparison is often made to justify the employment of the same, potentially improper, prior on $\dn{\beta}_0$ across models.
This usage is becoming standard, see for example \citep{Bayarri_etal_2012} and \citep{review}. It can be justified if, without essential loss of generality, we assume that the model has been parametrized in an orthogonal fashion, so that $\mt{X}_0^T\mt{X}_1 = \mt{0}$. 
In the special case where $M_0$ is the ``null" model, with only the intercept, this assumption can be justified, if we assume, again without loss of generality, that the columns of the design matrix of the full model have been centred on their corresponding means, which makes the covariates orthogonal to the intercept, and gives the intercept an interpretation that is ``common" to all models. Regarding the error variance, although it is also standard to be treated as a ``common" parameter across models, in this paper we follow the ``intrinsic prior methodology" (see for example \citep{moreno_giron_2008}) and we treat it as a model specific parameter. As we will see later in this Section, this causes no issues about the indeterminacy of Bayes factors due to the ``intrinsification". 

The sufficient statistic for $\dn{\beta}_1, \sigma_1^2$, under $M_1$, is:
$$
\dn{T}_2 = (\dn{T_{11}}^T, T_{12})^T = \Big( \, \widehat{\dn{\beta}\:}\!_1^T, \, \mbox{RSS}_1 \Big)^T, 
$$ 
where  $\widehat{\dn{\beta}\:}\!_1$ denotes the maximum likelihood estimate of $\dn{\beta}_1$ and $\mbox{RSS}_1$ the residual sum of square under model $M_1$. To simplify notation we drop the double index from $\dn{T_{11}}$ and $T_{12}$ and therefore we use the symbols $\dn{T_{1}}$ and $T_{2}$ to denote the sufficient statistics under model $M_1$ for $\dn{\beta}_1, \sigma_1^2$. 

Let $\dn{ y }^*=(y_1^*, \dots, y_{n^*}^*)^T$ be a training (imaginary) sample of size $n^*$: $k_1+1 \leq n^* \leq n$ and let $\mt{X}_0^*$ and $\mt{X}_1^*=\left[\mt{X_0}^* | \mt{X}_{e_1}^* \right]$ denote the corresponding imaginary design matrices. As before, we assume that all matrices are of full rank. Furthermore let $\mt{P}_0^* = \mt{X}_0^*\left({\mt{X}_0}^{*^T} \mt{X}_0^* \right)^{-1} {\mt{X}_0^*}^T$, $\mt{P}_1^* = \mt{X}_1^*\left({\mt{X}_1}^{*^T} \mt{X}_1^* \right)^{-1} {\mt{X}_1^*}^T$, 
$\widehat{\dn{b}\:}\!_{e_1}^* = D^{*^{-1}}\mt{X}_{e_1}^{*^T}(\mt{I}_{n^*}-\mt{P}^*_0)\dn{y}^*$ and $D^* = \mt{X}_{e_1}^{*^T}(\mt{I}_{n^*}-\mt{P}_0^*)\mt{X}_{e_1}^*$. 
If $(\dn{T_{1}}^{*^T}, T_{2}^*) = \Big( \, {\widehat{\dn{\beta}\:}\!_1^*}^{T}, \, \mbox{RSS}_1^* \Big)^T$ denotes the sufficient statistic under model $M_1$ using data $\dn{ y }^*$ and design matrix $\mt{X}_1^*$ (where for data $\dn{ y }^*$, $\widehat{\dn{\beta}\:}\!_1^*$ denotes the maximum likelihood estimate of $\dn{\beta}_1$ and $\mbox{RSS}_1^*$ the residual sum of square under model $M_1$), then $\dn{T_{1}^*}$ can be decomposed as follows:
$$
\dn{{T_1}}^* = \left({\widehat{\dn{b}\:}\!_0^*}^T, {\widehat{\dn{b}\:}\!_{e_1}^*\!}^{T} \right)^T = \left(\left(\widehat{\dn{\beta}\:}\!_0^*-\left( \mt{X}_0^{*^T} \mt{X}_0^* \right)^{-1} \mt{X}_0^{*^T} \mt{X}_{e_1}^* \widehat{\dn{b}\:}\!_{e_1}^* \right)^T, {\widehat{\dn{b}\:}\!_{e_1}^*\!}^{T} \right)^T
$$
with $\widehat{\dn{\beta}\:}\!_0^*$ being the maximum likelihood estimate, given data $\dn{ y }^*$, of $\dn{\beta}_0$. For the following analysis we use $M_0$ as the reference model.

Under the PEP methodology, using the power likelihood, we can prove that conditionally on $M_0$,  $\dn{T}_{1}^* \sim  N_{k_1} \left(\left(\dn{\beta}_0^T, \dn{0}^T\right)^T, \delta \sigma_0^2 \left(\mt{X}_1{^*}^T \mt{X}_1^* \right)^{-1} \right)$ and $T_{2}^* \sim  \mbox{Gamma} \left(\frac{n^*-k_1}{2}, \frac{1}{2\delta \sigma_0^2} \right)$. Thus
\vspace{-0.2cm}
\small
\begin{eqnarray*} 
m_0^N(\dn{t}_1^*, {t_2^*}|\delta) &\propto& |\mt{X}_{e_1}^{*T}(I_{n^*}-\mt{P}_0^*)\mt{X}_{e_1}^*|^{\frac{1}{2}}\left( 2 \pi \delta \right)^{-\frac{k_1-k_0}{2}}
\int \int \left( 2 \pi \delta \sigma^2_0 \right)^{-\frac{k_0}{2}} |\mt{X}_0{^*}^T \mt{X}_0^*|^{\frac{1}{2}}\\ && \exp\left[- \frac{1}{2 \delta \sigma_0^2} \left( \widehat{\dn{\beta}\:}\!_0^* - \dn{\beta}_0 \right)^T \left( \mt{X}_0{^*}^T \mt{X}_0^* \right) \left( \widehat{\dn{\beta}\:}\!_0^* - \dn{\beta}_0 \right) \right]\frac{\left(\frac{1}{2\delta\sigma_0^2} \right)^{\frac{n^*-k_1}{2}}}{\Gamma\left(\frac{n^*-k_1}{2} \right)}\\  && {t_2^*}^{\frac{n^*-k_1}{2}-1} \, \exp \left[ -\frac{t_2^*+ {\widehat{\dn{b}\:}\!_{e_1}^*\!}^{T}\mt{X}_{e_1}^{*^T}(\mt{I}_{n^*}-\mt{P}_0^*)\mt{X}_{e_1}^*\widehat{\dn{b}\:}\!_{e_1}^*}{2\delta\sigma_0^2}\right] \sigma_0^{-(k_1-k_0)} \sigma_0^{-(1+d_0)}  d\dn{\beta}_0 d\sigma_0 \nonumber\\
&\propto& 2^{-1} \pi^{\frac{k_0-k_1}{2}} (2\delta)^{\frac{d_0}{2}}|\mt{X}_{e_1}^{*T}(I_{n^*}-\mt{P}_0^*)\mt{X}_{e_1}^*|^{\frac{1}{2}}\frac{\Gamma\left(\frac{n^*-k_0+d_0}{2} \right)}{\Gamma\left(\frac{n^*-k_1}{2} \right)} \\ && \frac{{t_2^*}^{\frac{n^*-k_1}{2}-1}}{\left(t_2^*+ {\widehat{\dn{b}\:}\!_{e_1}^*\!}^{T}\mt{X}_{e_1}^{*^T}(\mt{I}_{n^*}-\mt{P}_0^*)\mt{X}_{e_1}^*\widehat{\dn{b}\:}\!_{e_1}^* \right)^{\frac{n^*-k_0+d_0}{2}}},
\end{eqnarray*}
\normalsize
while conditionally on $M_1$, $\dn{T}_{1}^* \sim  N_{k_1} \left(\dn{\beta}_1^T, \delta \sigma_1^2 \left(\mt{X}_1{^*}^T \mt{X}_1^* \right)^{-1} \right)$, $T_{2}^* \sim  \mbox{Gamma} \left(\frac{n^*-k_1}{2}, \frac{1}{2\delta \sigma_1^2} \right)$ and thus  
\small 
\begin{eqnarray*} 
m_1^N(\dn{t}_1^*, {t_2^*}|\delta) &\propto& \int \int \left( 2 \pi \delta \sigma^2_1 \right)^{-\frac{k_1}{2}} |\mt{X}_1{^*}^T \mt{X}_1^*|^{\frac{1}{2}} \, \exp\left[- \frac{1}{2 \delta \sigma_1^2} \left( \dn{t}_1^* - \dn{\beta}_1 \right)^T \left( \mt{X}_1{^*}^T \mt{X}_1^* \right) \left( \dn{t}_1^* - \dn{\beta}_1 \right) \right]\\  &&\frac{\left(\frac{1}{2\delta\sigma_1^2} \right)^{\frac{n^*-k_1}{2}}}{\Gamma\left(\frac{n^*-k_1}{2} \right)} \, {t_2^*}^{\frac{n^*-k_1}{2}-1} \, \exp \left[ -\frac{{t_2^*}}{2\delta\sigma_1^2}\right] \sigma_1^{-(1+d_1)} d\dn{\beta}_1 d\sigma_1 \nonumber\\
&\propto& 2^{-1} (2\delta)^{\frac{d_1}{2}} \frac{\Gamma\left(\frac{n^*+d_1-k_1}{2} \right)}{\Gamma\left(\frac{n^*-k_1}{2} \right)} {t_2^*}^{-\frac{d_1+2}{2}}.
\end{eqnarray*}
\normalsize

Using \citep{fouskakis_2019}, the PEP prior is

\begin{equation*} 
\pi_1^{PEP} ( \dn{\beta}_1, \sigma_1 ) = \pi^N_1( \dn{\beta}_1, \sigma_1) E_{{\dn{T}_{1}^*},{T_2^*}}^{M_1} \left[ \frac{m_0^N(\dn{t}_1^*, {t_2^*}|\delta)}{m_1^N(\dn{t}_1^*, {t_2^*}|\delta)}\right],
\end{equation*}
resulting in 
\begin{eqnarray*}
\pi_1^{PEP} ( \dn{\beta}_1, \sigma_1 ) &=& \sigma_1^{-(1+d_1)} \frac{\Gamma\left(\frac{n^*-k_0+d_0}{2} \right)}{\Gamma\left(\frac{n^*-k_1+d_1}{2} \right)} \pi ^{\frac{k_0-k_1}{2}}(2\delta)^{\frac{d_0-d_1}{2}}|\mt{X}_{e_1}^{*T}(I_{n^*}-\mt{P}_0^*)\mt{X}_{e_1}^*|^{\frac{1}{2}} \\ 
&&
\left( \delta \sigma_1^2\right)^{\frac{(d_1-d_0)+(k_0-k_1)}{2}} \int \int \frac{\left(\frac{{t_2^*}}{\delta\sigma_1^2}\right)^{\frac{n^*-k_1+d_1}{2}}}{\left(\frac{t_2^*+ {\widehat{\dn{b}\:}\!_{e_1}^*\!}^{T}\mt{X}_{e_1}^{*^T}(\mt{I}_{n^*}-\mt{P}_0^*)\mt{X}_{e_1}^*\widehat{\dn{b}\:}\!_{e_1}^* }{\delta\sigma_1^2}\right)^{\frac{n^*-k_0+d_0}{2}}} 
\end{eqnarray*}

\begin{eqnarray*}
&& \left( 2 \pi \delta \sigma^2_1 \right)^{-\frac{k_1}{2}} |\mt{X}_1{^*}^T \mt{X}_1^*|^{\frac{1}{2}} \, \exp\left[- \frac{1}{2 \delta \sigma_1^2} \left( \dn{t}_1^* - \dn{\beta}_1 \right)^T \left( \mt{X}_1{^*}^T \mt{X}_1^* \right) \left( \dn{t}_1^* - \dn{\beta}_1 \right) \right]\\  
&&\frac{\left(\frac{1}{2\delta\sigma_1^2} \right)^{\frac{n^*-k_1}{2}}}{\Gamma\left(\frac{n^*-k_1}{2} \right)} \, {t_2^*}^{\frac{n^*-k_1}{2}-1} \, \exp \left[ -\frac{{t_2^*}}{2\delta\sigma_1^2}\right]d{\dn{t}_1^*}d{t_2^*}.
\end{eqnarray*}
We know that, under $H_1$, $S=\frac{{t_2^*}}{\delta\sigma_1^2} \sim X^2_{n^*-k_1}$ and 
$$
Z^2=\left(\frac{{\widehat{\dn{b}\:}\!_{e_1}^*\!}^{T}\mt{X}_{e_1}^{*^T}(\mt{I}_{n^*}-\mt{P}_0^*)\mt{X}_{e_1}^*\widehat{\dn{b}\:}\!_{e_1}^* }{\delta\sigma_1^2}\right) \sim X^2_{k_1-k_0}\left( \lambda = \delta^{-1} \sigma_1^{-2} \dn{\beta}_{e_1}^{*^T}\mt{X}_{e_1}^{*^T}(\mt{I}_{n^*}-\mt{P}_0^*)\mt{X}_{e_1}^*\dn{\beta}_{e_1}^*\right),
$$ 
i.e. the non-central $X^2$ distribution with $k_1-k_0$ degrees of freedom and non-centrality parameter $\lambda$. Therefore, if we denote by
\begin{equation*}
A = \sigma_1^{-(1+d_1)} \frac{\Gamma\left(\frac{n^*-k_0+d_0}{2} \right)}{\Gamma\left(\frac{n^*-k_1+d_1}{2} \right)} \pi ^{\frac{k_0-k_1}{2}}(2\delta)^{\frac{d_0-d_1}{2}}|\mt{X}_{e_1}^{*T}(I_{n^*}-\mt{P}_0^*)\mt{X}_{e_1}^*|^{\frac{1}{2}}\left( \delta \sigma_1^2\right)^{\frac{(d_1-d_0)+(k_0-k_1)}{2}},
\end{equation*}
we have
\small
\begin{eqnarray*}
\pi_1^{PEP} ( \dn{\beta}_1, \sigma_1 ) &=& A \ E_{S,Z^2}^{M_1} \left[ \frac{S^{\frac{n^*-k_1+d_1}{2}}}{\left(S+Z^2\right)^\frac{n^*-k_0+d_0}{2}} \right]\\
&=& A \ \sum_{j=0}^{\infty} \frac{\left(\frac{\lambda}{2}\right)^j \exp(-\lambda/2)}{j!} \ E_{S,T}^{M_1}\left[ \frac{S^{\frac{n^*-k_1+d_1}{2}}}{\left(S+T\right)^\frac{n^*-k_0+d_0}{2}} \right]
\end{eqnarray*}
\normalsize
with $T\sim X^2_{k_1-k_0+2j}$.
Thus 
\small 
\begin{eqnarray*}
\pi_1^{PEP} ( \dn{\beta}_1, \sigma_1 ) &=& A \ 2^{\frac{(d_1-d_0)-(k_1-k_0)}{2}}\frac{\Gamma\left[\frac{2n^*-2k_1+d_1}{2}\right]}{\Gamma\left[\frac{n^*-k_1}{2}\right]} \sum_{j=0}^{\infty} \frac{\left(\frac{\lambda}{2}\right)^j \exp(-\lambda/2)}{j!} \frac{\Gamma\left[\frac{n^*+(d_1-d_0)-k_1+2j}{2}\right]}{\Gamma\left[\frac{2n^*+d_1-k_0-k_1+2j}{2}\right]}\\
&=& A \ 2^{\frac{(d_1-d_0)-(k_1-k_0)}{2}}\frac{\Gamma\left[\frac{2n^*-2k_1+d_1}{2}\right]}{\Gamma\left[\frac{n^*-k_1}{2}\right]} \frac{\Gamma\left[\frac{n^*+(d_1-d_0)-k_1}{2}\right]}{\Gamma\left[\frac{2n^*+d_1-k_0-k_1}{2}\right]} \\ && M\left(\frac{n^*+d_0-k_0}{2}, \frac{2n^*+d_1-k_0-k_1}{2}, -\frac{\lambda}{2} \right),
\end{eqnarray*}
\normalsize
with $M(\cdot,\cdot,\cdot)$ being the Kummer hypergeometric function.
Therefore 
\begin{eqnarray*}
\pi_1^{PEP} ( \dn{\beta}_1, \sigma_1 ) &=& \pi_1^{PEP} ( \dn{\beta}_0, \dn{\beta}_{e_1}, \sigma_1 ) \\ 
&=& A \ 2^{\frac{(d_1-d_0)-(k_1-k_0)}{2}}\frac{\Gamma\left[\frac{2n^*-2k_1+d_1}{2}\right]}{\Gamma\left[\frac{n^*-k_1}{2}\right]} \frac{1}{\Gamma\left[\frac{n^*+d_0-k_0}{2}\right]}\\ &&\int_0^1{\exp\left(-\frac{\lambda t}{2} \right) t^{\frac{n^*+d_0-k_0}{2}-1} (1-t)^{\frac{n^*+d_1-d_0-k_1}{2}-1}}dt\\
&=& \frac{\Gamma(\frac{n^*+d_0-k_1}{2})\Gamma(\frac{n^*+(d_1-d_0)-k_1}{2})}{\Gamma(\frac{n^*-k_1}{2})\Gamma(\frac{n^*-k_1+d_1}{2})}\sigma_1^{-(d_0+1)} \\ && \int_0^{1} f_N \left( \dn{\beta}_{e_1} ; \dn{0}, \tfrac{\delta \sigma_1^2}{ t } \mt{V}_{e_1} \right)   f_B \left( t ; \tfrac{n^*+d_0-k_1}{2}, \tfrac{n^*+d_1-d_0-k_1}{2} \right)  dt,
\end{eqnarray*}
where $\mt{V}^{-1}_{e_1}=\mt{X}_{e_1}^{*^T}(\mt{I}_{n^*}-\mt{P}_0^*)\mt{X}_{e_1}^*$.
From the above we see that, conditionally on $\left( \dn{\beta}_0, \sigma_1 \right)$, the PEP prior is a beta mixture of a multivariate normal prior and overall can be written  using the following hierarchical structure
\begin{eqnarray}
&& \dn{\beta}_{e_1} | t, \sigma_1,  \dn{\beta}_0 \sim N_{k_1-k_0} \left( \dn{0}, \tfrac{\delta \sigma_1^2}{ t } \mt{V}_{e_1} \right), ~ 
t | \sigma_1 \sim \mbox{Beta} \left( \tfrac{n^*+d_0-k_1}{2}, \tfrac{n^*+d_1-d_0-k_1}{2} \right) , \label{PEP_mixture} 
\label{inverse_beta_hyper}\\ 
&&
\left( \dn{\beta}_0, \sigma_1 \right) \sim \pi_1^{PEP} \left( \dn{\beta}_0, \sigma_1 \right) \propto ~ \sigma_1^{-(d_0+1)}. \nonumber 
\end{eqnarray}
The EPP is directly available for $\delta=1$. 

Therefore, to sum-up, the PEP priors (or EPPs for $\delta=1$) for comparing models $M_0$ and $M_1$ are
$$
\left\{\pi_0^{PEP}(\dn{\beta}_0, \sigma_0) = \pi_0^N(\dn{\beta}_0, \sigma_0), \pi_1^{PEP} ( \dn{\beta}_1, \sigma_1 ) \right\},
$$
with 
\begin{eqnarray*}
	\pi_1^{PEP} ( \dn{\beta}_1, \sigma_1 ) &=& \pi_1^{PEP} \left( \dn{\beta}_0, \sigma_1 \right) \int_0^{1} \pi_1^{PEP} \left(\dn{\beta}_{e_1} , t | \sigma_1,  \dn{\beta}_0 \right) dt\\
	&\propto& \sigma_1^{-(d_0+1) } \int_0^{1} f_N \left( \dn{\beta}_{e_1} ; \dn{0}, \tfrac{\delta \sigma_1^2}{ t } \mt{V}_{e_1} \right)   f_B \left( t ; \tfrac{n^*+d_0-k_1}{2}, \tfrac{n^*+d_1-d_0-k_1}{2} \right)  dt.  
\end{eqnarray*}
In the above expression, $\pi_1^{PEP}\left(\dn{\beta}_{e_1} , t | \sigma_1,  \dn{\beta}_0 \right) = \pi_1^{PEP}\left(\dn{\beta}_{e_1} | t, \sigma_1,  \dn{\beta}_0 \right) \pi_1^{PEP}\left(t \right) $ is proper and $\pi_1^{PEP} \left( \dn{\beta}_0, \sigma_1 \right) \propto \sigma_1^{-(d_0+1) }$; i.e. the reference prior for the baseline model $M_0$. 
Therefore there are no issues about the indeterminacy of the Bayes factor, when comparing model $M_0$ to $M_1$, since after the ``intrinsification'' the unknown constants of the imposed priors will be the same for the two competing models; see Appendix A for a detailed explanation.

Under the usual case where the  reference model $M_0$ is the null model (with only the intercept), we have that the prior variance-covariance matrix of the model coefficients is given by 
$\mt{V}_{e_1}=(\mt{Z}_{e_1}^{*^T}\mt{Z}_{e_1}^*)^{-1}$, where $\mt{Z}_{e_1}^*$ is the matrix of the centred (at the mean) imaginary covariates. 
In practice, when using the PEP prior with centred covariates and imaginary design matrices equal to actual ones \citea{fouskakis_et.al_2013}{as in}, 
then the induced approach results in a mixture of $g$-priors \cite{liang_etal_2008} with a different hyper-prior on $g=\delta/t$. 

 
\begin{table}[!ht]
	\caption{EPPs and PEP priors, under the alternative hypothesis, for the normal linear case, for any $n^*$ and for $n^*=k_1+1$ (minimal training sample size)}
	\label{momrnormal}
	\begin{center}
		\footnotesize
		\begin{tabular}{l||c@{~}|c@{~}||c@{~}||c@{~}||c@{~} ||c@{~}|c} 
			\hline 
			& \multicolumn{6}{l}{ 
				$\pi_1(\dn{\beta}_1, \sigma_1) \propto \sigma_1^{-(d_0+1)} \int_0^{1} f_N \left( \dn{\beta}_{e_1} ; \dn{0}, a \tfrac{\sigma_1^2}{ t } \mt{V}_{e_1} \right)   f_B \left( t ; b_1, b_2 \right)  dt$ } \\
			& \multicolumn{6}{l}{ 
				$V(\dn{\beta}_{e_1} |   \dn{\beta}_0, \sigma_1) =  a E[t^{-1}] \mt{V}_{e_1} \sigma_1^2$ } \\ 
			\hline 
			& \multicolumn{2}{c||}{ } &   $ $ &   & $ $ 
			& \multicolumn{2}{c}{ Approximate $V(\dn{\beta}_{e_1} |   \dn{\beta}_0, \sigma_1)$ }\\ 
			& \multicolumn{2}{c||}{$a$} &   $b_1$ & $b_2$ & $E[t^{-1}]$ 
			& \multicolumn{2}{c}{ {\footnotesize \it (for large $n^*$) }}\\ 
			&  EPP       & PEP &  EPP/PEP & EPP/PEP & EPP/PEP     & EPP       & PEP\\      
			\hline 
			For any $n^*$ &  $1$ & $\delta$ &  $\tfrac{n^*+d_0-k_1}{2}$ & $\tfrac{n^*+d_1-d_0-k_1}{2}$ 
			&  $\frac{2n^*+d_1-2k_1-2}{n^*+d_0-k_1-2}  $  
			&  $2 \mt{V}_{e_1}  \sigma_1^2$  & $2\delta \mt{V}_{e_1} \sigma_1^2$  \\ 
			Minimal $n^*$ & &&&&&\\ 
			($n^*=k_1+1$)  &  $1$ & $\delta$ &  $\tfrac{d_0+1}{2}$ & $\tfrac{d_1-d_0+1}{2}$ &  $\frac{d_1}{d_0-1}$  
			&  $ \frac{d_1}{d_0-1} \mt{V}_{e_1} \sigma_1^2$ & $\delta \frac{d_1}{d_0-1} \mt{V}_{e_1} \sigma_1^2$  \\ 
			\hline 
		\end{tabular}
		\normalsize 
	\end{center}
\end{table}

Table \ref{momrnormal} summarizes the EPP and PEP priors, under the alternative hypothesis, 
for minimal training sample size ($n^*=k_1+1$) as well as for any training sample size 
$n^* \in [k_1+1,n]$. Concerning the prior distribution of $\dn{\beta}_{e_1} |   \dn{\beta}_0, \sigma_1$ 
(after integrating out the hyper-parameter $t$), 
for large $n^*$, its corresponding  variance will be 
equivalent to the variance of a   $g$-prior with $g=2$ and $g=2\delta$ for the EPP and PEP prior, respectively. 
Clearly, the PEP prior is more dispersed accounting 
for information equivalent to $n^*/2\delta$ additional data points, 
while EPP will account for $n^*/2$ additional data-points.
When we consider the EPP, with the minimal training sample, that is $n^*=k_1+1$, then  
$V(\dn{\beta}_{e_1} |   \dn{\beta}_0, \sigma_1)$ is similar to the variance of a g-prior 
with $g = d_1/(d_0-1)$. This means that it can be defined only for choices 
of $d_0>1$. 
On the other hand, $V(\dn{\beta}_{e_1} |   \dn{\beta}_0, \sigma_1)$ can be defined without any problem 
when we consider any training sample of size $n^*> k_1-d_0+2$.   
Finally, under the PEP prior, the variance of $\dn{\beta}_{e_1} |   \dn{\beta}_0, \sigma_1$ is 
further multiplied by $\delta$ making larger the spread of the prior and overall the imposed prior less informative. For this reason, the corresponding posterior summaries will be more robust 
to the specific choices of $d_1$ and $d_0$, especially when $\delta=n^*=n$ and $n$ is large. 

\section{Comparison with Other Scale Normal Mixtures Priors and Properties} 
\label{comp}
Generally, a wide range of prior distributions for variable selection in regression can be written with the following form of a normal scale mixture distribution: 
\begin{eqnarray}
&& 
\pi_1(\dn{\beta}_{e_1}, \dn{\beta_0}, \sigma_1 )   = \sigma_1^{-(d_0+1) } \int _0 ^{+\infty} f_{N_{k_1-k_0}} \left( \dn{\beta}_{e_1} ; \, \dn{0}, \, g \sigma_1^2 \mt{\Sigma}_{e_1}  \right) \pi_1(g) dg,
\label{gen_mixture}  
\end{eqnarray}
where $f_{N_{d}} \left( \dn{y} ; \, \dn{\mu}, \,  \mt{\Sigma}  \right)$ denotes the density of the $d$-dimensional Normal distribution with mean $\dn{\mu}$
and covariance matrix $\mt{\Sigma}$, evaluated at $\dn{y}$ and $\pi_1 (g)$ denotes the prior distribution of the parameter $g$ under model $M_1$. 
Under the PEP prior, the hyper-prior $\pi_1(g)$ for $g$ is given by 
$$
g \sim SGBP \Big( a=\tfrac{n^*+d_0-k_1}{2}, \, b=\tfrac{n^*+d_1-d_0-k_1}{2}, \, p=1, \, q=\delta, \, s=\delta  \Big)
$$
where $SGBP$ stands for the
{\it shifted generalized beta prime distribution} with density 
\begin{equation}
f(g; a, b, p, q, s) = \frac{ p \left( \frac{g-s}{q} \right)^{bp-1} \left( 1 + \left( \frac{g-s}{q} \right)^{p} \right)^{-a-b} }{ q B(a,b) }, ~ g\ge s~. 
\label{sgbp}
\end{equation}
The beta prime distribution is a special case of \eqref{sgbp} with $p=q=1$ and $s=0$. Furthermore, the generalized beta prime distribution is a special case of \eqref{sgbp} with $s=0$. 
In our case, since $q=s=\delta$, the density of the hyper-prior for $g$ simplifies to 
\begin{equation}
f(g; a, b, \delta) \propto (g-\delta)^{b-1} g^{-a-b},~ g\ge \delta,   
\label{hyperprior_gbetaprime}
\end{equation}
where  $a=\tfrac{n^*+d_0-k_1}{2}$ and $b=\tfrac{n^*+d_1-d_0-k_1}{2}$. 

From the above expression, it is evident that the PEP prior implements an indirect averaging approach across all values of $g \ge \delta$. For the recommended setup \citea{fouskakis_et.al_2013}{see} of $\delta = n^* =n$, this might look quite dramatic at the first sight. But in practice, it is reasonable, under lack of prior information, to consider at most a value of $g$ that  will  correspond to one unit of information. Moreover, in such cases, the shrinkage $w$ given by $\tfrac{g}{g+1} = \tfrac{\delta}{\delta + t}$ (see Section \ref{shrinkage}) should approach the value of one, such that most of the posterior information comes from the data. 
In the case where the likelihood mass supports  values of $g$ lower than $\delta$, 
this means that the data do not have enough information in order to estimate sufficiently the model coefficients. 
An unrestricted prior for $g$ leads to greater shrinkage towards the prior mean of model coefficients $\dn{\beta}$. The truncation avoids over-shrinkage and the posterior of $g$ will be concentrated at the value of $\delta$ ensuring a minimum value of shrinkage towards the prior. 

\begin{table}[!p]
\caption{Mixing distributions of $g$ under different prior setups ($g \geq s)$}
\label{mixdistr}
\begin{center}
\footnotesize
\begin{tabular}{l||c@{~}||c@{~}||c@{~}||c@{~} ||c@{~}||c} 
\hline 
             &  &\multicolumn{5}{c}{Parameters of the SGBP distribution} \\ 
Prior & hyper-prior &  $a$ & $b$ & $p$ & $q$ & $s$ \\
\hline 
PEP &  \\
~~~(General)     & SGBP &  $\tfrac{n^*-k_1}{2}$ & $\tfrac{n^*-k_1}{2}$ & $1$ & $\delta$ & $\delta$\\
~~~(Recommended)     & SGBP &  $\tfrac{n-k_1}{2}$ & $\tfrac{n-k_1}{2}$ & $1$ & n & n\\
EPP & \\ 
~~~(General)     & SGBP &  $\tfrac{n^*-k_1}{2}$ & $\tfrac{n^*-k_1}{2}$ & $1$ & 1 & 1\\
~~~(Recommended)     & SGBP &  $\tfrac{1}{2}$ & $\tfrac{1}{2}$ & $1$ & 1 & 1\\
Intrinsic &   SGBP &  $\tfrac{1}{2}$ & $\tfrac{1}{2}$ & $1$ & $\tfrac{n}{k_1+1}$ & $\tfrac{n}{k_1+1}$\\
Robust  & \\
~~~(General)     &  SGBP      & $a_r$              & $1$                & $1$   & $\frac{b_r+n}{\rho_{1,r}^{-1}}$ & $\frac{b_r+n}{\rho_{1,r}^{-1}}-b_r$ \\ 
~~~(Recommended) &  SGBP      & $1/2$              & $1$                & $1$   & $\frac{n+1}{k_0+k_1}$ & $\frac{n+1}{k_0+k_1}-1$ \\ 
MG$^{*}$ & \\ 
~~~(General) &  Beta$^\prime$ &  $a_{mg}+1$ & $b_{mg}+1$ & $1$ & $1$  &0 \\
~~~(Recommended)&  Beta$^\prime$ &  $1/4$ & $\frac{n-q_{mg}-5}{2}+\frac{3}{4}$ & $1$ & $1$  &0 \\
Hyper-$g$ & \\ 
~~~(General)    &  Beta$^\prime$ &  $\frac{a_h}{2}-1$  & $1$                & $1$   & $1$ & $0$\\ 
~~~(Recommended)&  Beta$^\prime$ & $1/2$              & $1$                & $1$   & $1$ & $0$\\ 
Hyper-$g/n$ & \\ 
~~~(General)    &  GBP &  $\frac{a_h}{2}-1$  & $1$                & $1$   & $n$ & $0$\\ 
~~~(Recommended)&  GBP &  $1/2$              & $1$                & $1$   & $n$ & $0$\\ 
Benchmark$^{**}$ & \\ 
~~~(General)    &  Beta$^\prime$ &  $c_b$  & $c_b ~ max(n,p^2)$                & $1$   & $1$ & $0$\\ 
~~~(Recommended)&  Beta$^\prime$ &  $0.01$  & $0.01 ~ max(n,p^2)$                & $1$   & $1$ & $0$\\
\hline 
\multicolumn{7}{p{15cm}}{\footnotesize \it $^{*}$\citep{maruama_george_2011} prior but only for the case where $q_{mg}<n-1$; where $q_{mg}$ is the dimension of an orthogonal matrix which diagonalizes $\dn{X}^T\dn{X}$.    }\\
\multicolumn{7}{p{15cm}}{\footnotesize \it $^{**}$Under the Benchmark prior, $p = k-1$ denotes the total number of regressors.}\\
\multicolumn{7}{p{15cm}}{\footnotesize \it SGBP: Shifted generalized beta prime distribution.} \\ 
\multicolumn{7}{p{15cm}}{\footnotesize \it GBP: Generalized beta prime distribution.} \\ 
\multicolumn{7}{p{15cm}}{\footnotesize \it Beta$^\prime$: Beta prime distribution.} \\ 
\end{tabular}
\normalsize 
\end{center}
\end{table}

Most of the known priors used for variable selection assume that $\Sigma_{e_1}^{-1}=\mt{X}_{e_1}^{^T}(\mt{I}_{n}-\mt{P}_0)\mt{X}_{e_1}$ in \eqref{gen_mixture}.  
This is also the case for the PEP prior if we consider  $\mt{X}_{e_1}^*=\mt{X}_{e_1}$ as in \citep{fouskakis_et.al_2013}. 
Similarly, the benchmark prior \cite{ley_steel}, the robust prior \cite{Bayarri_etal_2012}, 
the hyper-g and hyper-$g/n$ priors \cite{liang_etal_2008} can be written as in \eqref{gen_mixture} with the hyper-prior for $g$ to be as in \eqref{sgbp}; details are provided in Table \ref{mixdistr}, under the usual choice of $d_1=d_0=0$ for simplicity reasons. Additionally, the EPP, as shown above (and also in \citep{womack_2014}) can be written as in \eqref{gen_mixture}, but using imaginary design matrices in $\Sigma_{e_1}^{-1}$, with number of rows usually equal to the minimal training sample ($n^*=k_1+1$). 
Also, the intrinsic prior of \citep{casella_moreno_2006a} can be viewed as an EPP. In their approach, as an approximation, using ideas from the arithmetic intrinsic Bayes factor approach, they used the original design matrix 
in $\Sigma_{e_1}^{-1}$, with all $n$ rows, using an additional multiplicator in the covariance matrix of the normal component in \eqref{gen_mixture} given by $\tfrac{n}{k_1+1}$; see for example \citep{womack_2014}. Therefore, this intrinsic prior can be viewed as a PEP prior, with (a) $\mt{X}_{e_1}^*=\mt{X}_{e_1}$; (b) $n^*=k_1+1$ (minimal training sample) and (c) a model dependent power parameter $\delta = \tfrac{n}{k_1+1}$; in the rest of the paper will call this prior intrinsic. 
Finally the prior by \citep{maruama_george_2011} is also closely related, where in the normal component in \eqref{gen_mixture} the rotated coordinates are used, while the Zellner and Siow prior \cite{Zellner_siow_80a} is as in \eqref{gen_mixture} with the hyper-prior for $g$ to be an inverted Gamma distribution with parameters $1/2$ and $n/2$.

\citep{Bayarri_etal_2012} developed criteria (\textit{desiderata}) to be satisfied by objective prior distributions for Bayesian model choice.
Obviously PEP prior satisfies the basic criterion (\emph{C1}). Furthermore \citep{fouskakis-ntzoufras-brazilian} proved that the PEP prior leads 
to a consistent model selection procedure (criterion \emph{C2}). \citep{fouskakis_ntzoufras_report_2016} showed that the PEP prior satisfies  the information consistency criterion (\emph{C3}). Additionally, as shown here, for $d_0=0$, the PEP prior belongs to a more general class of conditional
priors
\begin{equation}
\label{gc}
\pi_1(\dn{\beta}_{e_1}, \dn{\beta}_0, \sigma_1) \propto \sigma_1^{-1-(k_{1} -k_0)} h_1 \big( \tfrac{\dn{\beta}_{e_1}}{\sigma_1} \big),
\end{equation}
where $h_1(\cdot)$ is a proper density with support $\mathbb{R}^{k_{1} - k_0}$.
\citep{Bayarri_etal_2012} prove that the group invariance criterion (\emph{C7}) hold if and only if 
$\pi_1(\dn{\beta}_{e_1}, \dn{\beta}_0, \sigma_1)$ has the form of \eqref{gc}. Additionally, if $h_1(\cdot)$ is symmetric around zero, which is the case under the PEP prior, predictive matching
criterion (\emph{C5}) also holds. When, finally $\mt{X}_{e_1}^*=\mt{X}_{e_1}$, the conditional scale matrix has the form $\Sigma_{e_1}^{-1}=\mt{X}_{e_1}^{^T}(\mt{I}_{n}-\mt{P}_0)\mt{X}_{e_1}$ and then null predictive matching, dimensional predictive matching and the measurement invariance criterion (\emph{C6}) hold, according to \citep{Bayarri_etal_2012}.

\section{Posterior Distributions of Model Parameters} 

In this Section we the present posterior distributions of model parameters, under the PEP approach. For compatibility with the mixtures of g-prior, we work with the hyper-parameter $g=\delta/t$.

\subsection{Full Conditional Posteriors and Gibbs Sampling} \label{S4.1}
Under the PEP approach, the full conditional posterior distribution of $\dn{\beta}_{e_1}$ is a multivariate normal 
distribution of the form 
\begin{equation}
 \dn{\beta}_{e_1} | g, \sigma_1,  \dn{\beta}_0, \dn{y}, M_1 \sim N_{k_{e_1}}\left( \mt{W}_{e_1} \widetilde{\dn{\beta}\:}\!_{e_1},  \mt{W}_{e_1} ( \mt{X}_{e_1}^T \mt{X}_{e_1})^{-1} \sigma_1^2 \right)
  \label{full_betae} 
\end{equation}
where $k_{e_1}=k_1-k_0$ and 
\begin{eqnarray*}
\widetilde{\dn{\beta}\:}\!_{e_1} &=& (\mt{X}_{e_1}^T\mt{X}_{e_1})^{-1}\mt{X}_{e_1}^T( \dn{y} - \mt{X}_0^T \dn{\beta}_0 )  
                     =  \widehat{\dn{\beta}\:}\!_{e_1} - (\mt{X}_{e_1}^T\mt{X}_{e_1})^{-1}\mt{X}_{e_1}^T\mt{X}_0^T \dn{\beta}_0 \\ 
W_{e_1} & = &\left( w\mt{X}_{e_1}^T\mt{X}_{e_1} + (1-w) \mt{V}_{e_1}^{-1} \right)^{-1} \left( w\mt{X}_{e_1}^T\mt{X}_{e_1} \right) \\ 
w        & = & \frac{g}{g+1} = \frac{\delta}{\delta+t}; \\ 
&& \mbox{for~} \delta = 1  \Rightarrow \mbox{ EPP; } ~ \mbox{ ~for~} \delta>1 \Rightarrow \mbox{ PEP}.
\end{eqnarray*}
The matrix $W_{e_1}$ plays the role of a multivariate shrinkage factor which 
penalizes each coefficient locally, 
while $w$ is a global shrinkage factor which affects uniformly the posterior mean and posterior variance-covariance matrix. 
For example, if $w \rightarrow 0$ all the conditional posterior information is taken from the prior, 
while for  $w \rightarrow 1$  the conditional posterior information will be derived from the data.

Similarly we can obtain the full conditional posterior distributions of $\dn{\beta}_0$, $\sigma_1^2$ and $g$ by 
\begin{eqnarray}
 \dn{\beta}_0 | \dn{\beta}_{e_1}, \sigma_1,  g,  \dn{y}, M_1 &\sim& N_{k_0}\left( \widehat{\dn{\beta}\:}\!_0 -(\mt{X}_0^T\mt{X}_0)^{-1}\mt{X}_0^T\mt{X}_{e_1}^T \dn{\beta}_{e_1},  \, (\mt{X}_0^T\mt{X}_0)^{-1} \sigma_1^2 \right) \label{full_beta0}\\
 \sigma_1^2      | \dn{\beta}_0, \dn{\beta}_{e_1},  g,\dn{y}, M_1 &\sim&  IG \left(  \frac{n+k_{e_1}+d_0}{2}, \, \frac{ RSS_1 + \dn{\beta}_{e_1}^T \mt{V}_{e_1}^{-1} \dn{\beta}_{e_1}}{2}  \right) \label{fullpost_s2} \\
u           | \dn{\beta}_0, \dn{\beta}_{e_1}, \sigma_1, \dn{y}, M_1 &\sim&  CH\left( b, \, a+\frac{k_1-k_0}{2}+2, \, -\frac{1}{2\delta \sigma_1^2}\dn{\beta}_{e_1}^T \mt{V}_{e_1}^{-1} \dn{\beta}_{e_1} 
\right)  \nonumber \\
g &=& \delta /(1-u)\nonumber 
\end{eqnarray}
where $CH(p, q, s)$ is the confluent hypergeometric distribution with density function 
$$
f_{CH}( x; p, q, s ) \propto x^{p-1} (1-x)^{q-1} e^{-sx}~ \mbox{ for } 0\le x \le 1. 
$$
The above conditional distributions can be easily used to  implement a full Gibbs sampler in order to obtain any posterior estimates of interest for any specific model. 
Similarly, it can be used to build a Gibbs based variable selection sampler 
\citea{dellaportas_etal_02}{see for example in}
to obtain estimates of the posterior model weights.  
We can further simplify the Gibbs sampler by combining the posterior distributions of $\dn{\beta}_0$ and $\dn{\beta}_{e_1}$ given in \eqref{full_betae} and \eqref{full_beta0}. 
Finally, the full conditional posterior distribution of $\dn{\beta}_1^T = (\dn{\beta}_0^T, \dn{\beta}_{e_1}^T)$ is given by 
\begin{eqnarray*}
 \dn{\beta}_1 |  \sigma_1,  g,  \dn{y}, M_1 &\sim& N_{k_1}\left( \mt{W}_1 \widehat{\beta}_1,  \mt{W}_1 (\mt{X}_1^T \mt{X}_1)^{-1} \sigma_1^2 \right) \label{full_beta1} \\ 
 \mt{W}_1 &=& \left( w\mt{X}_1^T\mt{X}_1 + (1-w) \mt{T}_1  \right)^{-1}   w\mt{X}_1^T\mt{X}_1 \\  
 \mt{T}_1 &=&  \left(  
 \begin{array}{cc} 
 	\mt{0}_{k_0\times k_0} & \mt{0}_{k_0\times k_{e_1}} \\ 
 	\mt{0}_{k_{e_1}\times k_0} & \mt{V}_{e_1}^{-1} 
 \end{array} \right),  
\end{eqnarray*}
where $\mt{0}_{\ell_1 \times \ell_2}$ is a matrix of dimension $\ell_1 \times \ell_2$ with zeros, 
$w=g/(g+1)$ is the shrinkage parameter while $\widehat{\dn{\beta}\:}\!_1$ is the MLE for $\dn{\beta}_1$ of model $M_1$ given by 
$\widehat{\dn{\beta}\:}\!_1 =  (\mt{X}_1^T\mt{X}_1)^{-1}\mt{X}_1^T \dn{y}$.





\subsection{Marginal Likelihoods} \label{S4.2}
\label{marginal_likelihood}

The marginal likelihood conditionally on a value of $g$ is given by the usual marginal likelihood of the normal inverse gamma prior. 
Thus
\begin{equation} \label{marginal_lik_given_g}
f( \dn{y} | g, M_1) = C_1 \times  (g+1)^{\tfrac{n+d_0-k_1}{2}}  \left( 1+ g \, R_{10}\right)^{-\tfrac{n+d_0-k_0}{2}}, 
\end{equation}
with 
$R_{10}=\frac{1-R_1^2}{1-R_0^2}$; 
where $R_\ell^2$ is the coefficient of determination of model $M_\ell$ ($\ell \in \{0,1\}$), and $C_1$ being constant for all models (assuming that the covariates of $\mt{X}_0$ are included in all models) given by 
$$
C_1=2^{\frac{d_0}{2}-1}\pi^{\frac{k_0-n}{2}}  | \mt{X}_0^T\mt{X}_0 |^{-1/2}  \Gamma \left( \frac{n+d_0-k_0}{2} \right) 
(1-R_0^2)^{-\tfrac{n+d_0-k_0}{2}} || \dn{y} - \overline{y} \, \dn{1}_n ||^{-\frac{n+d_0-k_0}{2}}. 
$$


The full marginal likelihood $f( \dn{y} | M_1)$ is given by
\small 
\begin{eqnarray} 
f(\dn{y}|M_1) &=&   \tfrac{C_1}{\delta 
                B\left(  \tfrac{n^*+d_0-k_1}{2}, \, \tfrac{n^*+d_1-d_0-k_1}{2}\right)}   \nonumber \\ 
                && \times \delta^{1-b} \delta^{a+b}
                \int_\delta^{\infty} (1+g)^{\frac{n+d_0-k_1}{2}} (g-\delta)^{b-1}  g^{-a-b}
               \left( 1 + g R_{10} \right)^{-\frac{n+d_0-k_0}{2}} dg \nonumber \\ 
             &=&   \tfrac{C_1 }{\delta 
                B\left(  \tfrac{n^*+d_0-k_1}{2}, \, \tfrac{n^*+d_1-d_0-k_1}{2}\right)} 
               \times \delta^{1-b} \delta^{a+b} \delta^{-a} 
               (\delta+1)^{\frac{n+d_0-k_1}{2}}
               \left[ 1+ \delta R_{10} \right]^{-\tfrac{n+d_0-k_0}{2}} \nonumber \\ 
               && \times 
               \int_0^1 u^{b-1} (1-u)^{ \frac{k_1-k_0}{2}+a-1 } 
               \left(1-\tfrac{u}{\delta+1} \right)^{\frac{n+d_0-k_1}{2}}
               \left( 1 - u \tfrac{1-R_0^2}{1-R_0^2 + \delta(1-R_1^2)}\right)^{-\tfrac{n+d_0-k_0}{2}} du \nonumber \\ 
              &=& C_1 
               \times \frac{B\left(  \frac{k_1-k_0}{2}+a, \, b\right)}{ 
                B\left(  a, \, b\right)} 
               \times  
               (\delta+1)^{\frac{n+d_0-k_1}{2}}
               \left( 1+ \delta R_{10} \right)^{-\frac{n+d_0-k_0}{2}}
               \times \widetilde{F}_1(0) \label{marginal_like}
                \\ 
                && \mbox{with~}
                \widetilde{F}_1(0) = 
                F_1 \left( b, \, \tfrac{n+d_0-k_0}{2},\, -\tfrac{n+d_0-k_1}{2}, \,\tfrac{k_{e_1}}{2}+a+b; \,
                              \tfrac{1}{1+\delta R_{10}}, \,
                              \tfrac{1}{\delta+1}\right); \nonumber 
\end{eqnarray}
\normalsize 
where in the above $k_{e_1}=k_1-k_0$, $a=\tfrac{n^*+d_0-k_1}{2}$, $b=\tfrac{n^*+d_1-d_0-k_1}{2}$ (reminder from Eq. \ref{hyperprior_gbetaprime})  
and $F_1(a^\prime,b_1^\prime,b_2^\prime,c^\prime;x,y) $ is the hypergeometric function of two variables or Appell hypergeometric function given by 
$$
F_1(a^\prime,b_1^\prime,b_2^\prime,c^\prime;x,y) = 
\frac{1}{B(a^\prime,c^\prime-a^\prime)} \int_0^1 t^{a^\prime-1} (1-t)^{c^\prime-a^\prime-1}(1-xt)^{-b^\prime_1} (1-yt)^{-b^\prime_2} dt . 
$$ 
Note that the marginal likelihood of the reference model $M_0$ is   
$f(\dn{y}|M_0) = C_1$.

\subsection{Marginal Posterior Distribution of $g$} 

The marginal posterior distribution of $g$, under model $M_1$, is given by 
\be
\pi_1( g | \dn{y}) \equiv \pi( g | \dn{y}, M_1) = C_2 \times (1+g)^{\frac{n+d_0-k_1}{2}} \left( 1+ g \, R_{10}\right)^{-\frac{n+d_0-k_0}{2}}
	(g-\delta)^{b-1} g^{-a-b} 
\label{posterior_g}	
\ee
for $g\ge \delta$ with the normalizing constant $C_2$ given by 
\be
C_2 = \frac{\delta^a (\delta+1)^{-\tfrac{n+d_0-k_1}{2}} 
            \left( 1+ \delta \, R_{10}\right)^{\tfrac{n+d_0-k_0}{2}}
}
{B\left(  b, \, \tfrac{k_{e_1}}{2}+a \right)
 F_1 \left( b, \tfrac{n+d_0-k_0}{2}, -\tfrac{n+d_0-k_1}{2}, \tfrac{k_{e_1}}{2}+a+b; \tfrac{1}{1+\delta R_{10}}, \, \tfrac{1}{\delta+1}\right)}. 
\label{c2_constant}
\ee

The $\kappa$ posterior moment of $g$ is given by 
\begin{eqnarray*}
	E(g^\kappa|\dn{y}, M_1)   &=&   \delta^\kappa \frac{B \left( b, \, \tfrac{k_{e_1}}{2}+a-\kappa \right)}{B \left( b, \, \tfrac{k_{e_1}}{2}+a \right)} \frac{\breve{F}_1(\kappa)}{\breve{F}_1(0)} 
\end{eqnarray*}
where 
$$
\breve{F}_1(\kappa) = F_1 \left( b, \tfrac{n+d_0-k_0}{2}, -\tfrac{n+d_0-k_1}{2}, \tfrac{k_{e_1}}{2}+a+b-\kappa; \tfrac{1}{1+\delta R_{10}}, \, \tfrac{1}{\delta+1} \right)
$$
for $\kappa \in \{0,1,2, \dots\}$. Note that $\breve{F}_1(0)=\widetilde{F}_1(0)$.

The posterior expectation and variance of $g$ are now given by 
\begin{eqnarray*}
E(g|\dn{y}, M_1)   &=& \delta \, \frac{\widetilde{a} +b}{\widetilde{a}} \, \frac{\breve{F}_1(1)}{\breve{F}_1(0)} \\ 
V(g | \dn{y}, M_1) &=& \delta^2 \, \frac{ \widetilde{a}+b }{\widetilde{a}} \times  \frac{1}{\breve{F}_1(0)} \times 
\left( \frac{\widetilde{a}+b-1}{\widetilde{a}-1} \breve{F}_1(2) -\frac{\widetilde{a}+b}{\widetilde{a}} \frac{\breve{F}_1(1)^2}{\breve{F}_1(0)} \right) 
\end{eqnarray*}
where $\widetilde{a}= \tfrac{k_{e_1}}{2} +a-1$.

\section{Bayesian Inference of the Shrinkage Parameter} 
\label{shrinkage}
\subsection{Prior distribution of $w$}

Under model $M_1$, the imposed hyper-prior (or mixing) distribution for $w=\delta/(\delta + t)=g/(g+1)$ is induced via the 
beta hyper-prior for $t$ (see Eq. \ref{inverse_beta_hyper})
with parameters  given in Table \ref{momrnormal}. 
Hence, the resulted prior for $w$ is 
$$
w \sim BTPD \Big( a=\tfrac{n^*+d_0-k_1}{2}, \, b=\tfrac{n^*+d_1-d_0-k_1}{2}, \, \theta= \frac{\delta}{\delta+1} \, \lambda=1, \, \kappa=1  \Big)
$$
where $BTPD(a, b, \theta, \lambda, \kappa)$ is the {\it Beta truncated Pareto distribution} \cite{LDA2014} with parameters $a$, $b$, $\theta$, $\lambda$, $\kappa$ and density function  
$$
f( w;  a, b, \theta, \lambda, \kappa) = 
\frac{1}{B(a,b)} 
\frac{\kappa \theta^\kappa  w^{-\kappa-1} }{ 1 - \big(\tfrac{\theta}{\lambda}\big)^\kappa}  
\left[ 
	\frac{1-\big(\tfrac{\theta}{w}\big)^\kappa}
	     {1-\big(\tfrac{\theta}{\lambda}\big)^\kappa} \right]^{a-1}
\left[ 1-
\frac{1-\big(\tfrac{\theta}{w}\big)^\kappa}
{1-\big(\tfrac{\theta}{\lambda}\big)^\kappa} \right]^{b-1}
$$ 
for $\theta<w<\lambda$. 
The prior mean and variance of $w$ are now given by 
\begin{eqnarray*}
E(w) &=& ~_2F_1( 1, \, a, \, a+b; \, -1/\delta ) 
\mbox{ and } \\ 
Var(w)&=&~_2F_1( 2, \, a, \, a+b; \, -1/\delta )  - ~_2F_1( 1, \, a, \, a+b; \, -1/\delta ) ^2, 
\end{eqnarray*} 
where 
 $_2F_1(a_0,b_0, c_0; z)$ is the Gauss hyper-geometric function \cite{abramowitz_stegan_1970} given by 
$$
_2F_1(a_0,b_0, c_0; z) = \frac{1}{B(a_0,c_0-b_0)}\int_0^1 x^{b_0-1}(1-x)^{c_0-b_0-1}(1-zx)^{-a_0}dx~. 
$$ 

Equivalently we can show that the complementary shrinkage factor $u=1-w=1/(1+g)$ follows the truncated Compound Confluent Hypergeometric distribution; i.e. \small
$$
u \sim tCCH \left( t=a=\tfrac{n^*+d_0-k_1}{2}, \, q=b=\tfrac{n^*+d_1-d_0-k_1}{2}, \, s=0, \, r=a+b, \, s=0, \, 
v=\delta+1, \, \kappa= \tfrac{\delta+1}{\delta} \right),
$$
\normalsize with density expressed as
$$
f( u;  t, q, r, s, v, \kappa) = \frac{v^t \exp(s/v)}{B(t,q) \Phi_1(q,r,t+q,s/v,1-\kappa)} \frac{u^{t-1}(1-vu)^{q-1}\exp(-su)}{\left[\kappa+(1-\kappa)vu \right]^r}~~\mathbf{1}_{\{0<u<\frac{1}{v}\}},
$$ 
where $\Phi_1()$ is the Humbert series \cite{humbert}. Thus the PEP prior is a type of a ``{\it Confluent Hypergeometric Information Criterion}" (CHIC) $g$-prior introduced by \citep{li_clyde}. For a comparison with other CHIC $g$-priors, that have been appeared in literature, see \citep[Table 1]{li_clyde}.

In order to get more insight about the behavior of the prior distribution of $w$, 
we can obtain approximations of the prior mean and variance by using the first terms of a Taylor expansion given by 
\begin{equation}
\label{approx}
\mathbb{E}[ w(t) ] \approx w( \mu_t )  + \frac{1}{2} \frac{d^2w (\mu_t)}{dt^2} \sigma_t^2  \mbox{ and } 
\mathbb{V}[ w(t) ] \approx \left[ \frac{dw (\mu_t)}{dt} \right]^2 \sigma_t^2, 
\end{equation}
where $\mu_t$ and $\sigma_t^2$ are the prior mean and variance of the hyper-parameter $t$.  
By implementing the above approach, we obtain the approximations summarized in Table \ref{shrinkage_mean_sds_regression}; for the PEP prior we restrict attention on the choice of $\delta=n^*$. 
Note that for small training samples, the dimensions $k_0$ and $k_1$ may  influence the imposed prior, making it  sometimes more informative than intended. 

 \begin{table}[htb!]
 	\caption{Approximate prior means and variances of the shrinkage parameter $w$}
 	\label{shrinkage_mean_sds_regression}
 	\begin{center}
 		\footnotesize
 		\begin{tabular}{ll@{~}l@{~}||c@{~}|c@{~}||c@{~}||c@{~}} 
 			\hline 
 			&&   & \multicolumn{2}{c||}{ Prior Mean of $w$} & \multicolumn{2}{c}{Prior St. Deviation of $w$}  \\ 
 			&&   &  EPP       & PEP $(\delta=n^*)$  &  EPP &PEP $(\delta=n^*)$ \\			
 			\hline\hline 
 			\multicolumn{3}{l||}{ For large $n^*$}  &  $\frac{2n^*-2k_1+d_1}{3n^*-3k_1+d_1+d_0}$ & $1-\frac{1}{2n^*}$& $\frac{2}{9n^{*^{1/2}}}$  &  $\frac{1}{2{n^*}^{3/2}}$ \\ 
 			\hline
 			\multicolumn{3}{l||}{ Minimal $n^*$ ($n^*=k_1+1$)    }  &  &&& \\ 
 			&\multicolumn{2}{l||}{ Reference ($d_0=d_1=0$)} & $0.704$ & $1$ & $0.158$ & $ 1/\sqrt{8}k_1 $ \\
 			&               &           & {\footnotesize\it (for all $k_1$)} & {\footnotesize\it (for large $k_1$)} 
 			& {\footnotesize\it (for all $k_1$)} & {\footnotesize\it (for large $k_1$)} \\  
 			\cline{2-7}
 			&\multicolumn{2}{l||}{Jeffreys ($d_0=k_0$, $d_1=k_1$)} & & & & \\ 
 			&~~~&$k_0=1, k_1=2$             & $0.691$ & $0.86$ & $0.128 $      &0.071\\
 			\cline{3-7}
 			&							 &        &           &       &                                 & \\[-1em] 
 			&               & Large $k_0$, fixed $k=k_1-k_0$  & $1/2$   & $1$   & $\sqrt{\dfrac{k+1} {8k_0^2}}$ & $\sqrt{\dfrac{2(k+1)}{k_0^4}}$\\
 			&							 &        &           &       &                                 & \\[-1em] 
 			\cline{3-7}
 			&							 &        &           &       &                                 & \\[-1em] 
 			&               & Fixed $k_0$, large $k=k_1-k_0$  & $ 1 $     & $1$   & $\sqrt{\dfrac{2(k_0+1)}{k^2}}$  & $\sqrt{\dfrac{2(k_0+1)}{k^4}}$ \\[-1em]
 			&							 &        &           &       &                                 & \\
 			\hline 
 		\end{tabular} 
 		\normalsize 
 	\end{center}
 \end{table}

From Table \ref{shrinkage_mean_sds_regression} it is evident that when considering 
the usual EPP setup with the minimal training sample, then the prior mean of the shrinkage 
is far away from the value of one for specific cases (e.g. for the reference prior or for the Jeffreys' dependence prior when 
$k_0=1$ and $k_1=2$). 
This is not the case for the PEP prior for which the prior mean of the shrinkage is close to one even for 
models of small dimension; 
for example, under  the reference prior and for $k_0=1$ and $k_1=2$  we obtain a prior mean of the shrinkage equal to $0.86$ and 
a prior standard deviation of the shrinkage equal to $0.071$. 
Generally the global shrinkage parameter $w$ under the PEP prior is close to the value of one implying that the prior is generally 
non-informative since most of the information is taken from the data.

\subsection{Marginal Posterior Distribution of $w$} \label{S5.2}

Under model $M_1$, the marginal posterior distribution of the shrinkage parameter $w$ can directly derived by \eqref{posterior_g} resulting in 
\begin{eqnarray*} 
\pi_1(w | \dn{y}) \equiv \pi(w | \dn{y}, M_1) &=& C_2 \times (1+\delta)^{b-1} \\ 
&& \times (1-w)^{\tfrac{k_{e_1}}{2}+a-1} w^{-a-b} \left( w -\tfrac{\delta}{\delta+1} \right)^{b-1} \big\{ 1-w \, (1-R_{10})\big\}^{-\tfrac{n+d_0-k_0}{2}}, \\  
&& \mbox{for~} \tfrac{\delta}{\delta+1} \le w \le 1 
\end{eqnarray*} 
where the constant $C_2$ given by \eqref{c2_constant}. 

The posterior $\kappa$ moment is given by 
\begin{eqnarray}
E( w^\kappa | \dn{y}, M_1) &=& C_2 \times \left( 1+ \delta R_{10} \right)^{-\tfrac{n+d_0-k_0}{2}}  \delta^{-a+\kappa} (\delta+1)^{\tfrac{n+d_0-k1}{2}-\kappa} B\left( b, \tfrac{k_{e_1}}{2}+a \right) \widetilde{F}_1 (\kappa) \nonumber \\ 
&=& \left( \frac{\delta}{\delta+1} \right)^\kappa \times   \frac{\widetilde{F}_1 (\kappa)}{\widetilde{F}_1 (0)} 
\end{eqnarray}
where 
\small 
	$$
	\widetilde{F}_1(\kappa) = F_1 \left( b, \, \tfrac{n+d_0-k_0}{2},  \, -\tfrac{n+d_0-k_1}{2}+\kappa,  \, \tfrac{k_{e_1}}{2}+a+b; \,  \tfrac{1}{1+\delta R_{10} }, \, \tfrac{1}{\delta+1} \right)
	$$
\normalsize
for $\kappa \in \{0,1,2, \dots\}$.
Therefore the posterior expectation and variance of $w$ are directly derived as 
\begin{eqnarray}
E(w | \dn{y}, M_1) &=& \frac{\delta}{\delta+1} \times   \frac{\widetilde{F}_1 (1)}{\widetilde{F}_1 (0)} 
\label{post_mean_w} \\ 
V(w | \dn{y}, M_1) &=& \left( \frac{\delta}{\delta+1} \right)^2 
\frac{\widetilde{F}_1 (2)\widetilde{F}_1 (0)-\widetilde{F}_1 (1)^2}{\widetilde{F}_1 (0)^2}~.  
\nonumber
\end{eqnarray}

\section{Bayesian Model Averaging, Computation and Model Search}  

Details about the implementation of Bayesian model averaging (BMA) for PEP priors using the mixture representation are provided in Appendix B. 
Moreover, in Appendix C we present three alternative MCMC schemes for implementing model search, model averaging and computation of the parameters of interest under the PEP prior using the mixture representation. 
Specifically, we provide details about: 
(a) a vanilla $MC^3$ algorithm \cite{madigan_york_95}; 
(b) an $MC^3$ algorithm conditional on $g$; and 
(c) an MCMC variable selection scheme based on the Gibbs variable selection of \citep{dellaportas_etal_02}. 
All these three schemes are summarized in Algorithms 2--4 at Appendix C.

\section{Simulation Study}
\label{simulation}

In this Section we illustrate the proposed methodology in simulated data. 
We compare the performance of PEP prior and the intrinsic prior, the latest as presented in Section \ref{comp}. 
We consider 100 data sets of $n=50$ observations with $p=15$ covariates.
We run two different scenarios. Under Scenario 1 (independence) all covariates are generated from a multivariate Normal distribution with mean vector 
$\dn{ 0}$ and covariance matrix $\mt{ I }_{ 15 }$, while
the response is generated from
\begin{equation} \label{ss}
Y_i \sim N \big( 4 + 2 X_{ i1 } - X_{ i5 } + 1.5 X_{ i7 } + X_{ i, 11} +
0.5 X_{ i, 13 }, 2.5^2 \big), 
\end{equation}
for $i = 1, \dots, 50$. Under Scenario 2 (collinearity), the response is generated again from \eqref{ss}, but this time only the first 10 covariates are
generated from a multivariate Normal distribution with mean vector $\dn{ 0
}$ and covariance matrix $\mt{ I }_{ 10 }$, while
\begin{equation} \label{new3-1}
X_{ ij } \sim N \big( 0.3 X_{ i1 } + 0.5 X_{ i2 } + 0.7 X_{ i3 } + 0.9 X_{
	i4 } + 1.1 X_{ i5 }, 1 \big), 
\end{equation}
for $j = 11, \dots, 15; ~i = 1,
\dots, 50$.

With $p = 15$ covariates there are only 32,768 models to compare; we were
able to conduct a full enumeration of the model space, obviating the need
for a model-search algorithm in this example. 

Regarding the prior on model space we consider the uniform prior on model space (uni), as well as the uniform prior on model size (BB), as a special case of the beta-binomial prior \cite{scott_berger_2010}; thus in what follows we compare the following methods: PEP-BB, PEP-Uni, I-BB and I-Uni; the first two names denote the PEP prior under the uniform prior on model space and the uniform prior on model size respectively and the last two names the intrinsic prior under the uniform prior on model space and the uniform prior on model size respectively.

\begin{figure}[hb!]
	\centering{}
	\includegraphics[scale=0.3]{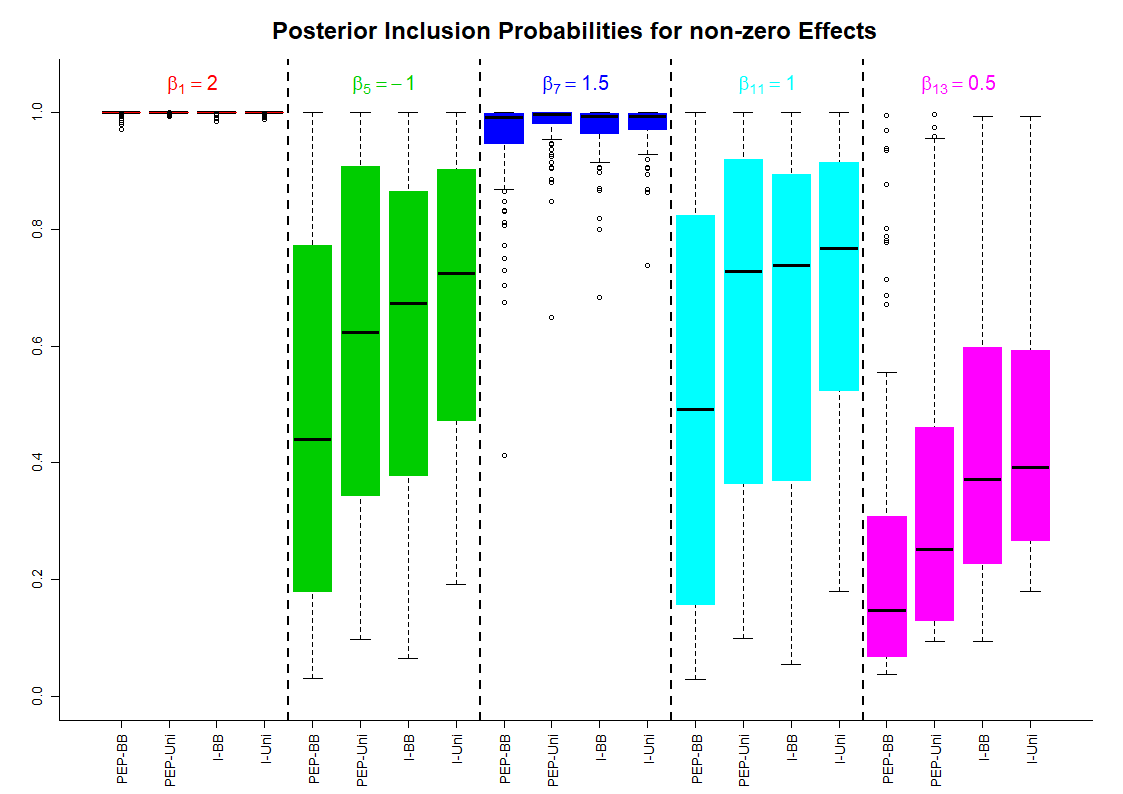}
	\caption{Simulation Scenario 1: Marginal Inclusion Probabilities for Non-Zero Effects.}
	\label{ex1_plot1}
\end{figure}

\begin{figure}[htp!]
	\centering{}
	\includegraphics[scale=0.3]{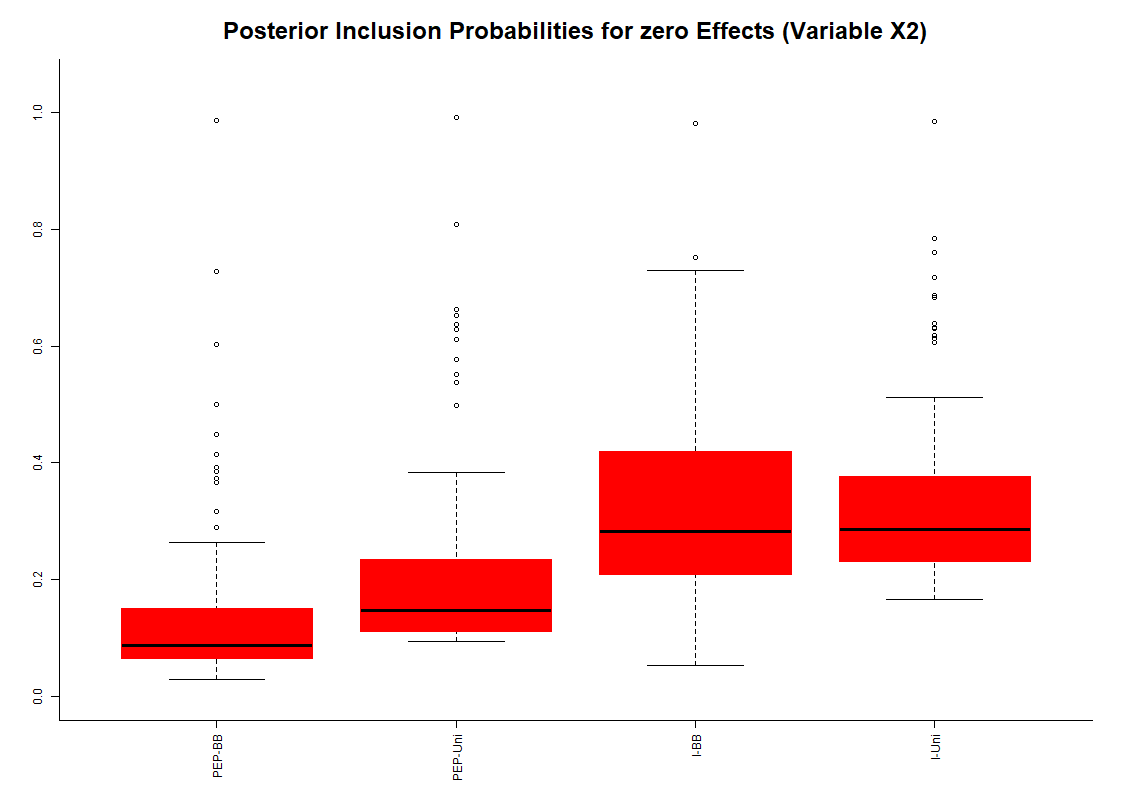}
	\caption{Simulation Scenario 1: Marginal Inclusion Probabilities for $X_2$ representing Covariates with Zero Effects.}
	\label{ex1_plot2}
\end{figure}

Under Scenario 1, the size of the posterior covariate inclusion probabilities for the non-zero effects (see Figure \ref{ex1_plot1}), for each method, is in agreement with the size each covariate's effect as expected. Hence the posterior inclusion probabilities for $X_1$ ($\beta_1=2$) are equal to one, with almost zero sampling variability, followed by $X_7$ ($\beta_7=1.5$) with posterior inclusion probabilities close to one, but with almost all values over 80\%. For covariates $X_5$ and $X_{11}$, the picture for their posterior inclusion probabilities is almost identical due to the same magnitude of the effects in absolute values (equal to one). 
Moreover, we observe large sampling variability within and across methods. Finally, all methods fail to identify $X_{13}$ ($\beta_{13}=0.5$) as an important covariate of the model, with the intrinsic approaches giving higher inclusion probabilities on average (around 40\%). Nevertheless, the posterior inclusion probabilities for $X_{13}$ are slightly higher on average and more dispersed across different samples, than the zero effects (see Figure \ref{ex1_plot2} for a representative example). 
Due to the independence of the covariates, we get similar results as the ones presented in Figure \ref{ex1_plot2} for the remaining zero effect covariates, and therefore plots are omitted for brevity reasons. Concerning the comparison of the different methods we observe that: (a) PEP is systematically more parsimonious than intrinsic, as previously reported in bibliography; (b) PEP-BB is more parsimonious than PEP-Uni; (c) I-BB supports slightly more parsimonious solutions than I-Uni. The last finding seems, at a first glance, surprising since the BB prior on model space is promoted in bibliography as a multiplicity adjustment prior. Nevertheless, in this example, the mean of the inclusion probabilities, under the uniform prior, in each data set is around 0.46, which is slightly reduced after the BB implementation to 0.44, leaving the results virtually unchanged.      
This is in accordance with what is expected by this prior, since it places a $U$ shaped distribution on the prior probabilities of each model depending on its dimensionality. 
This results in: 
(a) shrinkage of the inclusion probabilities when the observed proportion of variables and the average of posterior inclusion probabilities under the uniform prior is small (resulting in good sparsity properties in large $p$ problems); 
(b) inflation  of the inclusion probabilities when the observed proportion of variables and the average of posterior inclusion probabilities under the uniform prior is high (leading to posterior support of over-fitted models; a case which is largely neglected in the bibliography); and 
finally (c) leaving virtually unchanged the posterior inclusion probabilities when the observed proportion of variables and the average of posterior inclusion probabilities under the uniform prior  are close to $0.5$. 
The latter is the case here, where the number of true effects is $5$ out of $15$ (33\%) and the average of the posterior inclusion probabilities under the uniform prior on the model space is equal to $0.46$. 

\begin{figure}[htb!]
	\centering{}
	\includegraphics[scale=0.3]{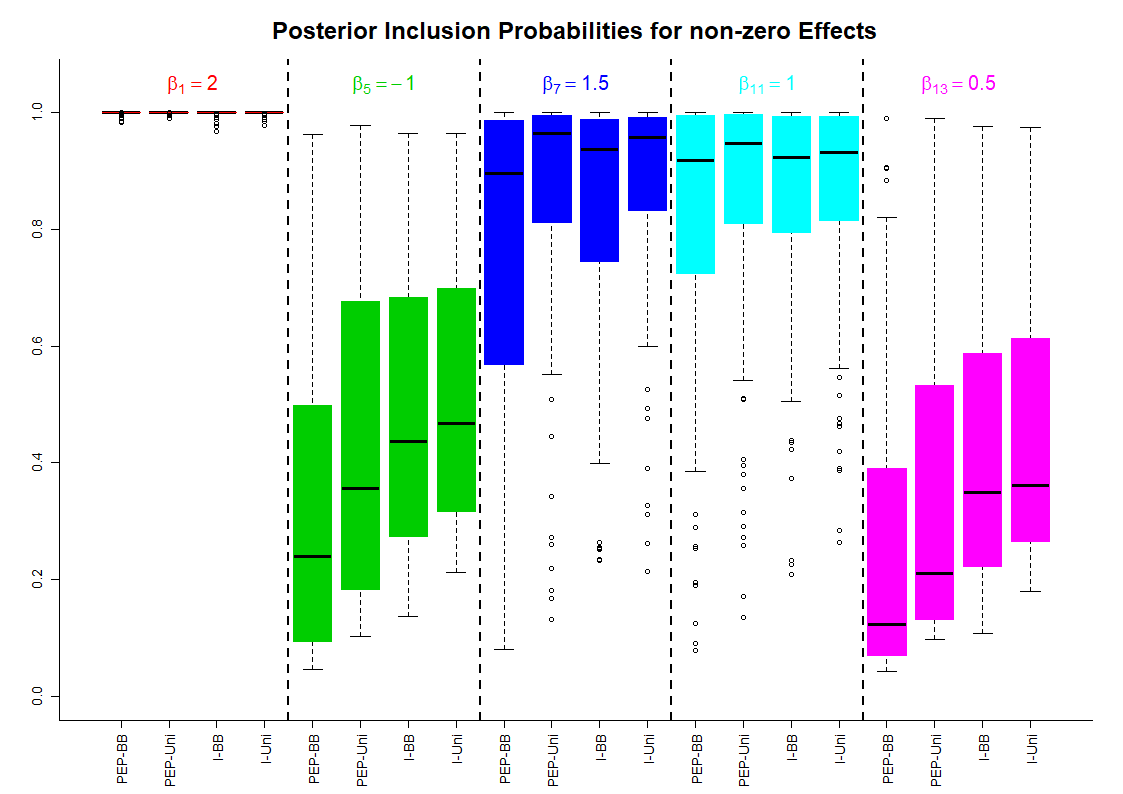}
	\caption{Simulation Scenario 2: Marginal Inclusion Probabilities for non-zero Effects.}
	\label{ex2_plot1}
\end{figure} 
 
\begin{figure}[htb!]
	\centering{}
	\includegraphics[scale=0.3]{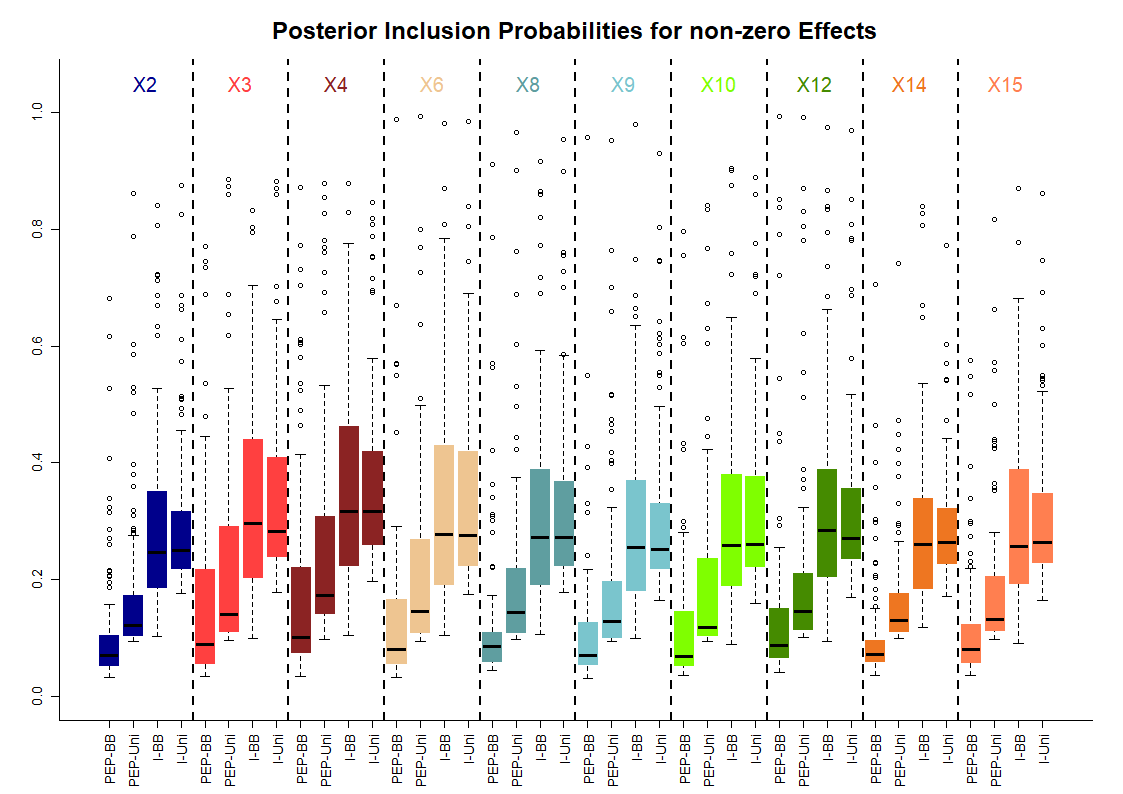}
	\caption{Simulation Scenario 2: Marginal Inclusion Probabilities for Covariates with Zero Effects.}
	\label{ex2_plot2}
\end{figure}  

Under Scenario 2, the posterior inclusion probabilities for $X_1$ and $X_{13}$ (see Figure \ref{ex2_plot1}), have similar picture as the ones in Scenario 1. For variable $X_7$ we observe again high inclusion probabilities, but this time with higher uncertainty. Due to collinearity, the posterior inclusion probabilities for covariates $X_5$ and $X_{11}$ are not longer similar, with the later being close to one (similar picture with the posterior inclusion probabilities for covariate $X_7$ with true effect equal to 1.5), while the first one ($X_5$) demonstrates posterior inclusion probabilities around 0.4, or lower, depending on the method. In Figure \ref{ex2_plot2}, the posterior inclusion probabilities for all the zero effects are presented. In all cases those are lower under the PEP prior, with the PEP-BB to behave the best. Moreover, they are differences across covariates, depending on collinearities, mainly in the variability across samples.  
 
\section {SDM Dataset} \label{SDM}

In this Section we consider the SDM data that contains $p = 67$ potential regressors for
modelling annual GDP growth per
capita between 1960 and 1996 for $n = 88$ countries. More details about this dataset can be found in \citep{ley_steel}.

We compare the performance of the PEP prior (with $\delta=n^*=n$) with that of other scale normal mixtures priors, as presented in Table \ref{mixdistr}. Specifically, we implement the methods based on:
(a) the PEP prior with $\delta=n^*=n$;
(b) the intrinsic prior \cite{womack_2014};
(c) the hyper-$g$ and the hyper-$g/n$ priors \cite{liang_etal_2008} with $a_h=3$;
(d) the robust prior \cite{Bayarri_etal_2012} with $a_r=0.5$;
(e) the benchmark prior \cite{ley_steel} with $c_b=0.01$. 
For comparative reasons we have also included the $g$-prior \cite{zellner_76} with $g=n$.

All methods have been implemented in \texttt{MultiBUGS}  and \texttt{R2MultiBUGS} \cite{multibugs}, using the Gibbs variable selection sampler \cite{dellaportas_etal_02}; see Appendix C for details (Algorithm 4). 
The obtained results have been generated using 100K MCMC iterations and a 10K burnin period and were additionally compared and validated using an $MC^3$ based algorithm, as described in Appendix C, for the PEP and the intrinsic priors, the \texttt{BAS} package in \texttt{R} \cite{BAS} for the $g$, hyper-$g$ and hyper-$g/n$ priors and the \texttt{BayesVarSel} package in \texttt{R} \cite{bayesvarsel} for the robust prior.

Regarding the prior on model space we consider, as before, the uniform prior on model space (uni), as well as the uniform prior on model size (BB), as a special case of the beta-binomial prior \cite{scott_berger_2010}. 

Posterior variable inclusion probabilities and posterior distributions of the model dimension across all visited models are presented in Figures 1--2, respectively, at Appendix D, along with detailed discussion and comparison of the related results obtained by all competing methods under consideration using the two prior distributions on the model space. 

\begin{figure}[htb!]
	\centering{}
	\includegraphics[scale=0.3]{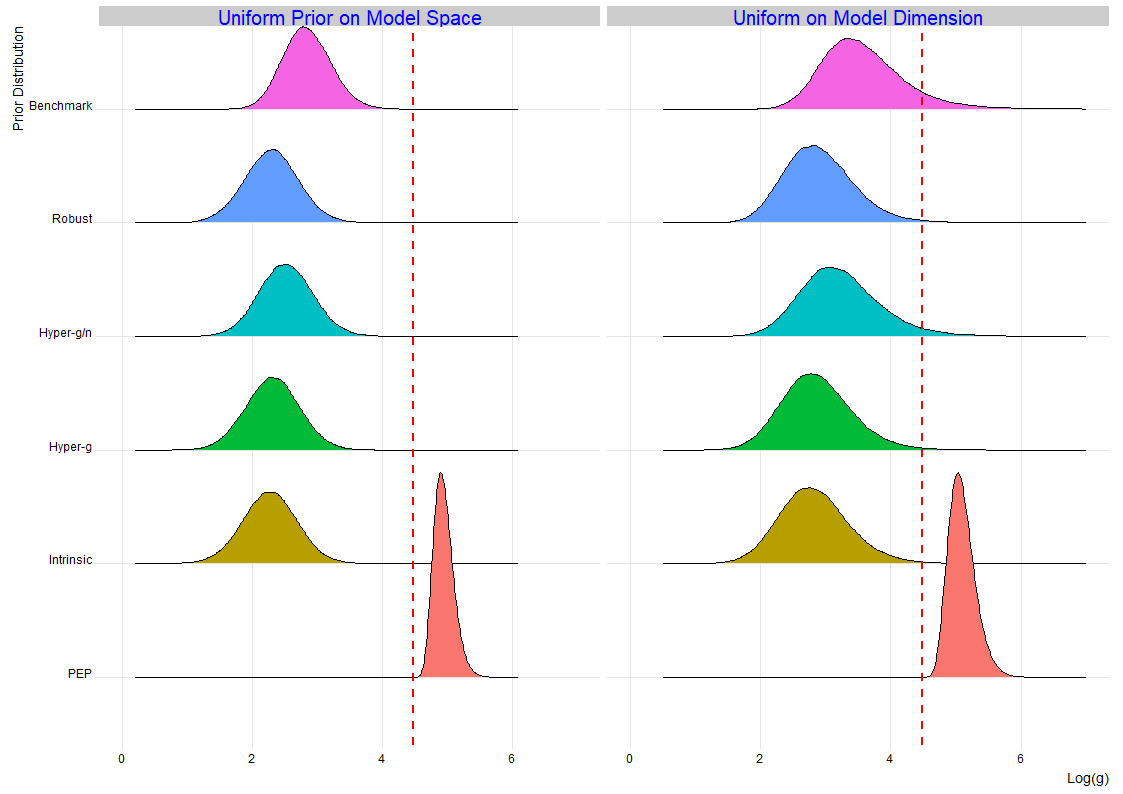}
	\vspace{-1.5em} 
	\caption{SDM Dataset: Posterior densities of $log(g)$ for each method (100K MCMC iterations).}
	\label{sdm_boxplot_logg}
\end{figure}

An interesting point of discussion is the fact that the lower bound imposed on $g$ seems to drive the final results, under the PEP prior. For the recommended PEP prior specification ($\delta=n$), the posterior density function of $g$ is zero if $g<\delta=n$. Of course we can still specify the PEP prior with smaller values of $\delta$ in order to consider different weighting of the imaginary data.  
By this way, the bound (via the choice of $\delta$) can be lower and thus we might leave $g$ to take values in a wider range.
In Figure  
\ref{sdm_boxplot_logg} we present the posterior densities of $log(g)$ for each competing method (except the $g$-prior, where $g$ is fixed and equal to $n$), under both priors on model space,  based on the MCMC runs.
The vertical dashed lines are referring to the vertical line of $x=log(n)$. We do not observe any noticeable differences when we move from the uniform prior on the model space to the uniform prior on model size; in the latter case the posterior densities, under each competing method, have a slightly larger variance. On the other hand, there are differences on the posterior distributions of 
$log(g)$ when applying different priors on the model parameters. The posterior distributions of $log(g)$ for all priors, except for the ones under the PEP prior, look similar, with the ones under the hyper-$g/n$ and the benchmark priors to be slightly shifted to the right. Under the PEP prior the posterior distributions of $log(g)$ are centred on larger values and have smaller standard deviations. Still, the mode of the posterior distribution of $log(g)$, under the PEP prior, is away from the lower bound, while the posterior standard deviations are high enough to allow for a satisfactorily posterior uncertainty for $g$.
On the other hand, using other hyper-priors for $g$ without restricting the range of values, like the hyper-$g$ and hyper-$g/n$, results on high posterior standard deviations of $g$, which in combination with low posterior modes, may result in a ``waste'' of valuable posterior probability in informative prior choices (within each model) and to the inflation of the posterior probability of irrelevant models with low practical usefulness. This behaviour has two side effects: (a) the posterior probability of the MAP model is considerably lower than the one obtained by methods with fixed prior choices for $g$, and 
(b) the posterior inclusion probabilities for the non-important covariates will be inflated towards 0.5; see \citep{dellaportas_etal_2012} for an empirical illustration within the hyper-$g$ setup. 

\begin{figure}[htb!]
	\centering{}
	\includegraphics[scale=0.33]{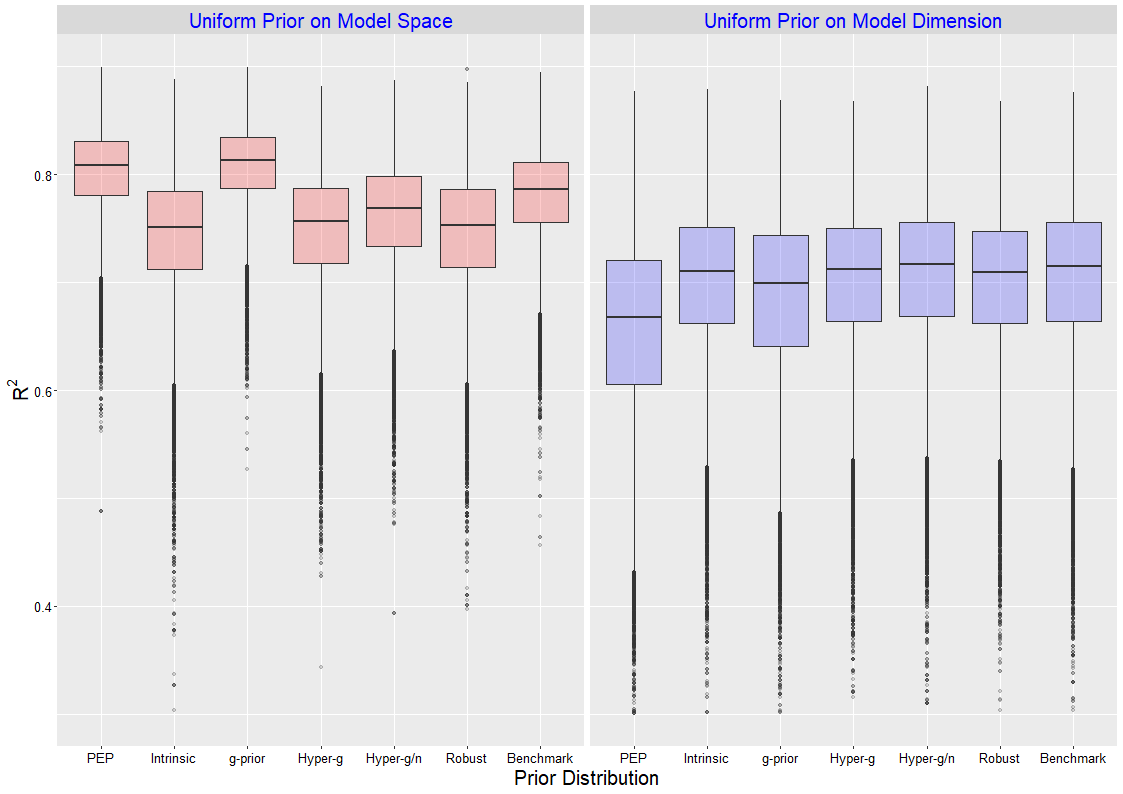}
	\vspace{-1.5em} 
	\caption{SDM Dataset: BMA posterior boxplots of $R^{2,(t)}$, $t=1,\dots,T$, for each method ($T=100K$ MCMC iterations).}
	\label{sdm_boxplot_R2}
\end{figure}

\begin{figure}[htb!]
	\centering{}
	\includegraphics[scale=0.33]{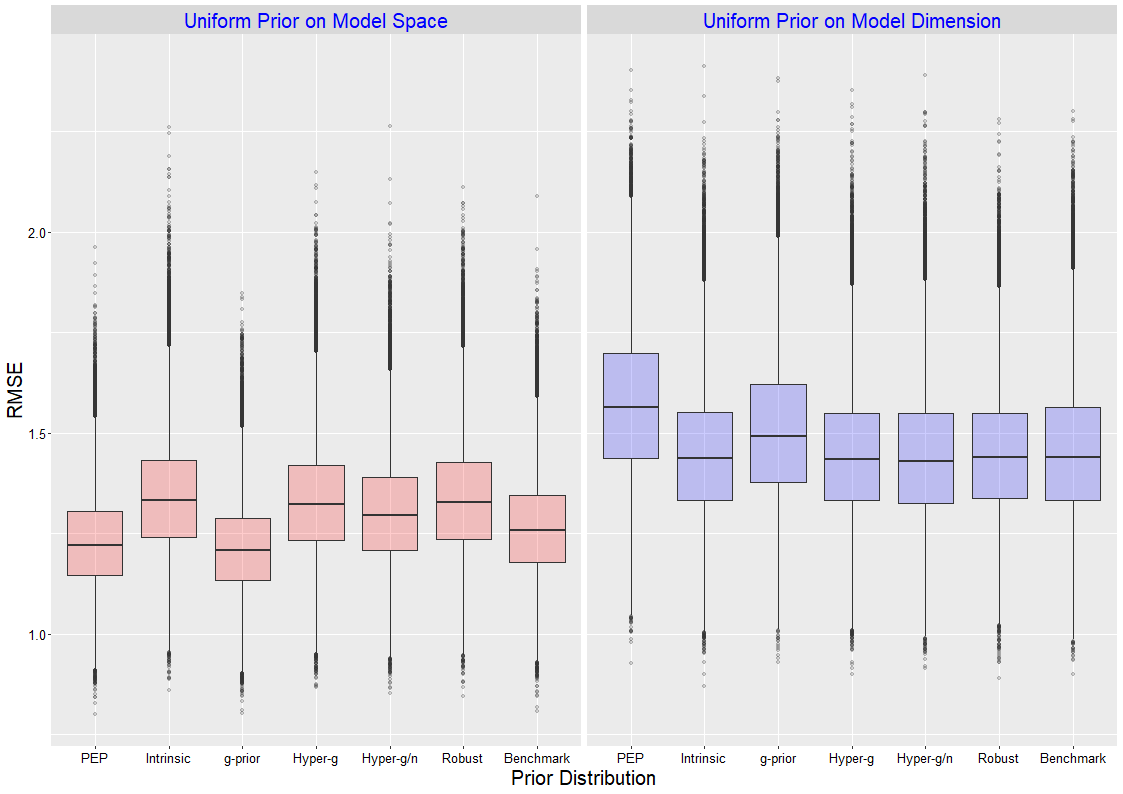}
	\vspace{-1.5em} 
	\caption{SDM Dataset:  BMA posterior boxplots of $RMSE^{(t)}$, $t=1,\dots, T$, for each method ($T=100K$ MCMC iterations).}
	\label{sdm_boxplot_RMSE}
\end{figure}   

In the following, we further focus on the implementation of a selection of measures concerning the fit and the predictive ability of the models. 
Figure \ref{sdm_boxplot_R2} presents boxplots of the Bayesian version of the coefficient of determination, for each model $M_\ell$, given by 
$R^2_\ell = 1 - \sigma^2_\ell/S_y^2$, 
where $S_y^2$ is the unbiased sample variance, while Figure \ref{sdm_boxplot_RMSE} presents boxplots of 
the root mean square error of the predictive values, of each model $M_\ell$, given by 
$RMSE_\ell=\frac{1}{n} \sum_{i=1}^n (Y_i - Y_{i,M_\ell}^{pred}),$ where $Y_{i,M_\ell}^{pred}$ were generated from the predictive distribution of each model $M_\ell$. 
These quantities have been calculated for each iteration $t$ of our Gibbs based variable selection algorithm.  Therefore, Figures \ref{sdm_boxplot_R2} and \ref{sdm_boxplot_RMSE} present the BMA posterior distribution of these quantities; see also Tables 1--2, at Appendix D, respectively, for the BMA point estimate of these quantities based on the posterior mean. 
Both of these quantities evaluate the in-sample overall fit of the models supported by each method. 
Overall we observe that the PEP and the $g$--prior under the uniform prior on model space achieve higher posterior $R^2$ values. On the other hand, all methods when combined with the beta-binomial (uniform on model dimension) prior support models with lower $R^2$ values. It is notable, that under the BB setup, the PEP prior supports models with lower $R^2$ values compared to other competing methods, which suggests that the BB prior in combination with PEP over-penalize the inclusion of covariates in sparse problems. 
Similar is the picture for the RMSE with the inverted relations between methods since lower RMSE values indicate better fitted models. Moreover, the variances of RMSE are now higher since the results are based on the predictive distribution rather than the posterior distribution of the error variance parameter.

Additionally we calculate the \textit{BMA--log predictive score} (BMA--LPS); see for example \citep{fernandez_etal_2001}. Specifically, we perform $\kappa$--fold cross--validation with $\kappa=8$, placing 11 randomly selected observations in each fold. We select $\kappa-1=7$ of the folds to form the modelling subsample and the remaining fold to form the validation subsample. We denote by $\mathbb{ M } = \{\dn{y^\mathbb{ M }}, \mt{X}^\mathbb{ M } \}$ the modelling subsample, of size $n^\mathbb{ M } = 77$, and by $\mathbb{ V } = \{\dn{y^\mathbb{ V }}, \mt{X}^\mathbb{ V } \} $ the validation subsample, of size $n^\mathbb{ V } = 11$, where $n=n^\mathbb{ M }+n^\mathbb{ V }$. The BMA--LPS then is given by
\begin{equation} \label{lps}
\mbox{BMA--LPS} = - \frac{1}{n^\mathbb{ V}} \sum_{i=1}^{n^\mathbb{ V}} \log f(y^\mathbb{ V}_i |  \dn{y^\mathbb{ M }}, \mt{X}^\mathbb{ V }),
\end{equation}
where
\begin{equation} 
f(y^\mathbb{ V}_i | \dn{y^\mathbb{ M }}, \mt{X}^\mathbb{ V }) 
= \frac{\sum_{M_\ell \in {\cal M}} f(y^\mathbb{ V}_i, \dn{y^\mathbb{ M }} |   M_\ell) \pi(M_\ell)}{\sum_{M_\ell \in {\cal M}} f(\dn{y^\mathbb{ M }} |  M_\ell) \pi(M_\ell)},
\label{ppo}
\end{equation}
with $\pi(M_\ell)$ denoting the prior probabilities of model $M_\ell$. Smaller values of BMA-LPS indicate better performance.

In this section we estimate $f(y^\mathbb{ V}_i |  \dn{y^\mathbb{ M }}, \mt{X}^\mathbb{ V })$ from the output of the Gibbs variable selection sampler (see Algorithm 4 at Appendix C) using as an estimate 
the posterior mean of the normal density evaluated at the model parameter values over the visited model at iteration $t$; for $t=1, \dots, T$. This automatically averages across all visited models. 

\begin{table}[htb!]
	\caption{SDM Dataset: Mean (Standard Deviations) over 8-fold CV of BMA estimates of log-predictive scores obtained using Gibbs based variable selection samplers}
	\label{logpredscore}
	\begin{center}
		\footnotesize
		\vspace{-2em} 
		\begin{tabular}{l@{~}c@{~}c@{~}c@{~}c@{~}c@{~}c@{~}c@{~}} 
			\hline 
			&  PEP  &Intrinsic &$g$-prior& Hyper-$g$& Hyper-$g/n$  & Robust & Benchmark \\ 
			\hline                      
			Uniform on models    &1.89 (0.31) &1.95 (0.71) &1.75 (0.40) & 1.69 (0.25) & 1.70 (0.27) & 1.70 (0.14) & 1.94 (0.35)\\
			Uniform on dimension &1.75 (0.37) &1.70 (0.31) &1.84 (0.45) & 1.72 (0.31) & 1.82 (0.31) & 1.75 (0.28) & 1.80 (0.40)\\
			\hline 
		\end{tabular} 
		\normalsize 
	\end{center}
\end{table}
\vspace{-1cm}

Table \ref{logpredscore} presents the means and the standard deviations of the BMA log-predictive score over the eight different modelling and validation combinations of subsamples (folds); 
see also Figure 3 at Appendix D for boxplots of these BMA predictive log-scores. 
BMA based on hyper-$g$, hyper-$g/n$ and the robust priors outperform the rest of the methods under the uniform prior on model space for this example. 
The rest of the methods ($g$-prior, PEP, intrinsic and the benchmark prior) have higher means and medians (Figure 3 at Appendix D) of log-predictive scores under the uniform prior on model space. 
Nevertheless, standard deviations do not support important differences between the log-predictive scores of all methods under comparison; note that the log-predictive score for the robust prior has much smaller variability than the corresponding scores for the rest of the methods while for the intrinsic prior the corresponding standard deviation is much higher due to the presence of one outlier. 
The picture is different when the uniform prior on model dimension is used. 
In this case, the intrinsic prior seems to perform better than the rest of the methods, followed by the hyper-$g$ prior while the PEP and the robust priors are tied performing slightly worse on average than the two previously mentioned methods. 
Note that, in terms of medians, the PEP prior performs better than all methods under consideration when the uniform prior on the model dimension is used; 
see Figure 3 at Appendix D.  
The rest of the methods ($g$-prior, hyper-$g/n$ and robust)  perform slightly worse  in terms of mean log-predictive score. 
The variability of the log-predictive scores is similar for all methods and does not   clearly suggests one of the methods as the best.



\section{Discussion}

In this article we have shown that the power-expected posterior prior, a generalization of the expected-posterior prior in objective Bayesian model selection, under normal linear models can be represented as a mixture of $g$-prior. This has the great advantage of being able to derive posterior distributions and marginal likelihoods in closed form, permitting fast calculations even when exploring high-dimensional model spaces. 

Our results imply that the PEP prior is more parsimonious than its competitors. 
We do not claim that this property is always the best practice in variable selection problems. The choice of parsimony or sparsity depends on the problem at hand. When we have a sparse dataset, where the important covariates are very few, then the PEP prior will act probably in a better way than other competitors, which may spend a big portion of the posterior probability to models that are impractical in terms of dimension and sparsity.

Additional future extensions of our method include the introduction of two different power parameters in order to derive a family of prior distributions, with members all the prior distributions for variable selection in regression that are written as mixtures of $g$-priors, that can be derived using either fixed or random imaginary data.  Furthermore, we plan to extent the applicability of PEP prior in cases where $k > n$. This can be done by (a) using shrinkage type of baseline priors, such as Lasso or Ridge; (b) assigning zero prior probability to models with dimension larger than $n$; and (c) mimicking formal approaches to use $g$-priors in situations where $k > n$, such as \citep{maruama_george_2011}, based on different ways of generalizing the notion of inverse matrices.

%

\bibliographystyle{agsm}

\bibliography{biblio3}

\end{document}


\normalem
\renewcommand{\baselinestretch}{1.55}\normalsize

\title{\vspace*{-0.5in}
\textbf{Electronic Appendix} \\ 
\textbf{\small of the paper entitled} \\
\textbf{\large ``Power-Expected-Posterior Priors as Mixtures of $g$-Priors in Normal Linear Models"}}

\author{
D.~Fouskakis\thanks{D.~Fouskakis is with the Department of
Mathematics, National Technical University of Athens, Greece; email
\texttt{fouskakis@math.ntua.gr}} \
and  I.~Ntzoufras\thanks{I.~Ntzoufras is with the Department of
Statistics, Athens University of Economics and Business, Greece; email
\texttt{ntzoufras@aueb.gr}} 
}

\date{}

\maketitle

\noindent

\section{Derivation of PEP Bayes Factors Under Improper Baseline Priors} 
\label{cancelation_of_constants} 

For the PEP prior setup (Eq. 3--6 in the main paper) with $\dn{\theta}_\ell = ( \dn{\beta}_\ell, \sigma^2_\ell )$, 
the Bayes factor for comparing model $M_1$ to $M_0$   takes the form
\begin{eqnarray}
B_{10} &=& \frac{ \int \int f(\dn{y} | \dn{\beta}_1, \sigma_1,  M_1 ) \pi_1^{PEP} ( \dn{\beta}_1, \sigma_1  ) d\dn{\beta}_1 d \sigma_1 }
{ \int \int f(\dn{y} | \dn{\beta}_0, \sigma_0,  M_0 ) \pi_0^{PEP} ( \dn{\beta}_0, \sigma_0  ) d\dn{\beta}_0 d \sigma_0 } \nonumber \\ 
&=& \frac{ \int \int f(\dn{y} | \dn{\beta}_1, \sigma_1, M_1 ) \left[ \int \frac{f(\dn{y}^* | \dn{\beta}_1, \sigma_1, \delta, M_1 ) \pi_1^N(\dn{\beta}_1, \sigma_1)}{m_1^N( \dn{y}^*| \delta)} m_0^N( \dn{y}^* | \delta) d \dn{y}^* \right] d\dn{\beta}_1 d \sigma_1 }
{ \int \int f(\dn{y} | \dn{\beta}_0, \sigma_0,  M_0 )  \pi_0^{N} ( \dn{\beta}_0, \sigma_0  ) d\dn{\beta}_0 d \sigma_0 } \label{eqA} 		
\end{eqnarray}
where, for $\ell = 0, 1$,
\begin{eqnarray*} 
	m_\ell^N( \dn{y}^*| \delta) 
	&=& \int \int f(\dn{y}^* | \dn{\beta}_\ell, \sigma_\ell, \delta, M_\ell ) \pi_\ell^N(\dn{\beta}_\ell, \sigma_\ell) d\dn{\beta}_\ell d\sigma_\ell \\ 
	&=& c_\ell \int \int f(\dn{y}^* | \dn{\beta}_\ell, \sigma_\ell, \delta, M_\ell ) \pi_\ell^U(\dn{\beta}_\ell, \sigma_\ell) d\dn{\beta}_\ell d\sigma_\ell. \\ 
	&=& c_\ell m_\ell^U(\dn{\beta}_\ell, \sigma_\ell). 
\end{eqnarray*}
Therefore, returning back to \eqref{eqA}, we have that  
\small
\begin{eqnarray} 
B_{10} 
&=& \frac{ \int \int \int f(\dn{y} | \dn{\beta}_1, \sigma_1, M_1 ) f(\dn{y}^* | \dn{\beta}_1, \sigma_1, \delta, M_1 ) c_1 \pi_1^U(\dn{\beta}_1, \sigma_1) \frac{ c_0 m_0^U( \dn{y}^* | \delta)}{c_1 m_1^U( \dn{y}^*| \delta)} d \dn{y}^*  d\dn{\beta}_1 d \sigma_1 }
{ \int \int f(\dn{y} | \dn{\beta}_0, \sigma_0,  M_0 )  c_0 \pi_0^{U} ( \dn{\beta}_0, \sigma_0  ) d\dn{\beta}_0 d \sigma_0 }. 	
\label{eqB}
\end{eqnarray}
\normalsize
As it is obvious from \eqref{eqB}, both normalizing constants, $c_0$ and $c_1$, cancel out.

\section{Bayesian Model Averaging Estimates}
\label{sec_bma} 

Let us consider a set of models $M_\ell \in {\cal M}$, with design matrices $\mt{X}_\ell$, where the covariates of the design matrix $\mt{X}_0$ (of the reference model $M_0$) are included in all models. Thus we assume as before that $\mt{X}_\ell = \left[\mt{X_0} | \mt{X}_{e_\ell} \right]$ and $\dn{\beta}_\ell = \left(\dn{\beta}_0^T, \dn{\beta}_{e_\ell}^T \right)^T$. We are interested in quantifying the 
uncertainty about the inclusion or exclusion of additional columns/covariates $\mt{X}_{e_\ell}$ of model $M_\ell$. 
Under this setup,  for any model $M_\ell \in {\cal M}$, given a set of new values of the explanatory variables
$\mt{X}_\ell^{new}= \left[ \mt{X}_0^{new}| \mt{X}_{e_\ell}^{new} \right]$, we are interested in estimating the corresponding posterior predictive distribution $f(\dn{y}^{new}|\dn{y}, \mt{X}_\ell^{new}, M_\ell)$.
Following \citep{liang_etal_2008}, for each model $M_\ell$, we consider the Bayesian model averaging (BMA) prediction point estimator which is the optimal one under squared error loss and is given by
\begin{eqnarray*} 
	\widehat{\dn{y}}^{new}_{BMA} 
	&=&E(\dn{y}^{new}|\dn{y}, \mt{X}^{new})  
	=\sum _{M_\ell \in {\cal M}} E(\dn{y}^{new}|\dn{y}, \mt{X}_\ell^{new}, M_\ell  ) \pi(M_\ell|\dn{y})   \\ 
	&=& \sum _{M_\ell \in {\cal M}} \mt{X}_\ell^{new} E( \dn{\beta}_\ell  | \dn{y},  M_\ell)  \pi(M_\ell|\dn{y})   \\ 
	&=& \sum _{M_\ell \in {\cal M}} \Big\{ \mt{X}_0^{new}     E(\dn{\beta}_0 | \dn{y},  M_\ell) 
	+\mt{X}_{e_\ell}^{new}E( \dn{\beta}_{e_\ell} | \dn{y},  M_\ell)\Big\} \pi(M_\ell|\dn{y}), 
\end{eqnarray*}   
where $\mt{X}^{new}$ is the given set of new values of all explanatory variables. 
Detailed derivations of the posterior means $E(\dn{\beta}_0 | \dn{y},  M_\ell) $ 
and $E( \dn{\beta}_{e_\ell} | \dn{y},  M_\ell)$ are provided in  Section 4.1 of the main paper. If we now further assume that $\mt{X}_0^T\mt{X}_\ell = \mt{0}$, then the posterior means of the coefficients are considerably simplified to 
\begin{eqnarray*} 
	E(\dn{\beta}_0 | \dn{y},  M_\ell)       &=& \widehat{\dn{\beta}\:}\!_0 
	\mbox{~and~} 
	E( \dn{\beta}_{e_\ell} | \dn{y}, M_\ell) = E\left(  \tfrac{g}{g+1} \Big| \dn{y}, M_\ell\right)  \widehat{\dn{\beta}\:}\!_{e_\ell},
\end{eqnarray*} 
where 
$\widehat{\dn{\beta}\:}\!_0 = (\mt{X}_0^T\mt{X}_0)^{-1}\mt{X}_0^T\dn{y}$ and 
$\widehat{\dn{\beta}\:}\!_{e_\ell} =(\mt{X}_{e_\ell}^T\mt{X}_{e_\ell})^{-1}\mt{X}_{e_\ell}^T\dn{y}$. 
Hence, assuming in a similar fashion that ${(\mt{X}_0^{new})}^T\mt{X}_\ell^{new} = \mt{0}$, 
the posterior predictive mean, under model $M_\ell$, is now reduced to 
\begin{eqnarray} 
\widehat{\dn{y}}_{M_\ell}^{new}                                                    
&=& E( \dn{y}^{new}  | \dn{y}, \mt{X}_\ell^{new}, M_\ell)
= \mt{X}_0^{new} \widehat{\dn{\beta}\:}\!_0 + \mt{X}_{e_\ell}^{new} E\left(  \tfrac{g}{g+1} \Big| \dn{y}, M_\ell\right) \widehat{\dn{\beta}\:}\!_{e_\ell} 
\label{ypredm} 
\end{eqnarray}
and the corresponding BMA point prediction estimate is now given by 
\begin{eqnarray} 
\widehat{\dn{y}}^{new}_{BMA}                                                  
&=& \mt{X}_0^{new} \widehat{\dn{\beta}\:}\!_0 + \sum _{M_\ell \in {\cal M}} \mt{X}_{e_\ell}^{new} E\left(  \tfrac{g}{g+1} \Big| \dn{y}, M_\ell\right) \widehat{\dn{\beta}\:}\!_{e_\ell}  \pi(M_\ell|\dn{y}). 
\label{bma} 
\end{eqnarray} 
The expected value of the posterior distribution of $w=g/(g+1)$ in \eqref{bma} is given in Section 5.2 of the main paper, 
while the posterior model probabilities $\pi(M_\ell | \dn{y}) \propto f(\dn{y}| M_\ell ) \pi(M_\ell)$ in \eqref{bma} can be calculated using the closed form expression of the marginal likelihood given in Section 4.2 of the main paper. 

For more details and computational strategies about the implementation of BMA we refer the reader to \citep{bma_pep}. 

\section{Computation and Model Search Algorithms} 
\label{cmsa}

The implementation of the variable selection methods for normal regression models based on mixtures of $g$-priors, including the PEP and the intrinsic priors, with mixture representation described in this article, can be easily implemented using full enumeration of the model space when the number of covariates is limited (for example, lower than 30) since the Bayes factors and many other posterior quantities of interest are given in closed form expressions.  

For large model spaces, several MCMC based algorithms, or the Bayesian adaptive sampling (BAS) introduced by \citep{clyde_etal_2011}, can be used for model search. 
Here we describe three alternative MCMC schemes for implementation in large spaces: 
\begin{enumerate} 
	\item $MC^3$ algorithm \cite{madigan_york_95}; see Algorithm \ref{algo1}. 
	
	\item $MC^3$ algorithm conditional on $g$; see Algorithm \ref{algo2}. 
	
	\item Fully MCMC Bayesian variable selection based on the Gibbs variable selection of \citep{dellaportas_etal_02}; see Algorithm \ref{algo3}.
\end{enumerate} 			

All these algorithms include a Bayesian variable selection (BVS) step of each covariate which is summarized in Algorithm \ref{algo0}.

The first MCMC scheme (Algorithm \ref{algo1}) is using directly the marginal likelihood which is readily available in (16) in the main paper. 
It might not be efficient in large spaces  due to the inversions of matrices. Moreover, in order to calculate this marginal likelihood we need to evaluate the Appell hypergeometric function and this evaluation might be unstable in some occasions. 
In such cases, \citep{liang_etal_2008} proposed to approximate the integral, which indirectly appears in the marginal likelihood via the the Appell hypergeometric function, by a simple Laplace approximation. 

Alternatively, we can easily avoid the evaluation of this function by implementing the second MCMC scheme (Algorithm \ref{algo2}); we now generate $g$ from the conditional posterior distribution; see (17) in the main paper, and  we decide about the inclusion or exclusion of each variable by using the marginal likelihoods given $g$; see Eq. 15 in the main paper.  

Finally, the last scheme (Algorithm \ref{algo3}) is a full MCMC scheme based on the Gibbs variable selection sampler of \citep{dellaportas_etal_02}. 
BMA estimates of any quantities are readily available using the MCMC output of this scheme. 
It is notable that this scheme  can be even implemented in ready-to-use MCMC software such as WinBUGS/OpenBUGS \cite{winbugs,openbugs}; see \citep{perrakis_ntzoufras_2018} for a similar treatment in WinBUGS/OpenBUGS. 

\label{appA}

\begin{algorithm}
	\colorbox{gray!25}{\parbox{0.92\textwidth}{
			\SetKwInOut{Input}{Input}
			\SetKwInOut{Output}{Output}
			\SetKwBlock{blocknotext}{~}{~}
			\SetKwBlock{step}{Step}{~}
			\SetKwBlock{mcmcmcstep}{Description of a BVS Step}{~}
			\SetKwBlock{ini}{Initialization:}{~}
			\SetKwBlock{schemeb}{SA version 1: Full SA (Scheme B)}{End-Full-SA}
			\SetKwBlock{schemec}{SA version 2: Partial SA (Scheme C)}{End-Partial-SA}
			\SetKwFor{For}{for}{~}{endfor}
			\SetKwFor{ForAll}{for~all}{~}{endfor}
			\SetKwFor{ForEach}{for~each}{~}{endfor}

			\mcmcmcstep{
				\blocknotext(\textbf{(a)} For a covariate $X_j$ propose to change its status:){
					--- Set $M_1$ the model which corresponds to $\dn{\gamma}^{(t-1)}$ \\ 
					--- Set $\dn{\gamma}^{prop}=\dn{\gamma}^{(t-1)}$ \\ 
					--- Propose to change the status of $j$ covariate by setting $\gamma^{prop}_j=1-\gamma^{(t-1)}_j$ \\ 
					--- Set $M_2$ the model which corresponds to $\dn{\gamma}^{prop}$ 
				}
				\vspace{-4ex}
				\blocknotext(\textbf{(b)} Model Update:){
					Accept the proposed model with probability 
					\vspace{-0.5em} 
					$$
					\alpha = \min\{1,A_j\}, 
					$$ 
					\vspace{-0.5em} 
					where $A_j$ is the ratio of the appropriate conditional/marginal posterior distributions of $M_2$ versus $M_1$ for covariate $X_j$. 
				}
				\vspace{-1em}
			} 
			\vspace{-4ex}
			\caption{Bayesian variable selection (BVS) step for each covariate $X_j$}
			\label{algo0}
		}
	}
\end{algorithm}

\begin{algorithm}
	\colorbox{gray!25}{\parbox{0.94\textwidth}{
			\SetKwInOut{Input}{Input}
			\SetKwInOut{Output}{Output}
			\SetKwBlock{blocknotext}{~}{~}
			\SetKwBlock{step}{Step}{~}
			\SetKwBlock{mcmcmcstep}{Description of a BVS Step}{~}
			\SetKwBlock{ini}{Initialization}{~}
			\SetKwBlock{schemeb}{SA version 1: Full SA (Scheme B)}{End-Full-SA}
			\SetKwBlock{schemec}{SA version 2: Partial SA (Scheme C)}{End-Partial-SA}
			\SetKwFor{For}{for}{~}{endfor}
			\SetKwFor{ForAll}{for~all}{~}{endfor}
			\SetKwFor{ForEach}{for~each}{~}{endfor}
			\SetKwInOut{notation}{Notation}
			
			\ini({\bf 1:} Substitute model indicator $M$ by variable inclusion indicators $\dn{\gamma}$.){
				Each model $M$ is identified by the usual vector of binary variable inclusion indicators $\dn{\gamma} = \big(\gamma_1, \dots \gamma_p \big)$; where $p = k-1$ denotes the total number of covariates under consideration.  }	
			\vspace{-4ex}
			\ini({\bf 2:} Set the starting model $M^{(0)}$:){Set the initial values of $\dn{\gamma}^{(0)}=\big( \gamma_1^{(0)}, \dots,  \gamma_p^{(0)}\big)$. Proposed starting model is the full one with $\gamma_j^{(0)}=1$ for all $j=1,\dots,p$.}
			
			\vspace{-4ex}
			\notation{$M_1$: current model in $t$ iteration; $M_2$: proposed model in $t$ iteration; $M_0$: reference model of PEP prior}

			\Output{Random sample $\dn{\gamma}^{(t)}$ for $t=1,\dots T$.} 
			 
			\For{$t=1$ \KwTo $T$}{
				\For{$j=1$ \KwTo $p$}{
					\blocknotext(BVS step for covariate $X_j$:){
						Impement BVS step of Algorithm \ref{algo0} with acceptance probability 
						$$
						\alpha = \min \left\{ 1, \frac{f(\dn{y}|M_2) \pi(M_2)}{f(\dn{y}|M_1) \pi(M_1)} \right\} 
						$$				    		
						with the marginal likelihoods $f(\dn{y}|M)$ given by (16) in the main paper and $\pi(M)$ being the prior probability of model $M$; $M \in \{ M_1, M_2\}$. 
					}
					\vspace{-2em} 
				}	
			}
			\blocknotext(END of the algorithm){}
			\vspace{-2em}
			\caption{Model search using $MC^3$ algorithm}
			\label{algo1}
		}
	}
\end{algorithm}

\begin{algorithm}
	\colorbox{gray!25}{\parbox{0.92\textwidth}{
			\SetKwInOut{notation}{Notation}			
			\SetKwInOut{Input}{Input}
			\SetKwInOut{Output}{Output}
			\SetKwBlock{blocknotext}{~}{~}
			\SetKwBlock{step}{Step}{~}
			\SetKwBlock{mcmcmcstep}{Description of an $MC^3$ Step}{~}
			\SetKwBlock{ini}{Initialization}{~}
			\SetKwBlock{schemeb}{SA version 1: Full SA (Scheme B)}{End-Full-SA}
			\SetKwBlock{schemec}{SA version 2: Partial SA (Scheme C)}{End-Partial-SA}
			\SetKwFor{For}{for}{~}{endfor}
			\SetKwFor{ForAll}{for~all}{~}{endfor}
			\SetKwFor{ForEach}{for~each}{~}{endfor}
			\ini({\bf 1:} Use variable inclusion indicators $\dn{\gamma}$ as in Algorithm \ref{algo1}){~}	
			\vspace{-5ex}
			\ini({\bf 2:} Set initial values of $g$ and $\dn{\gamma}$:){Set $g^{(0)} = n$ and
				$\dn{\gamma}^{(0)}=\big( \gamma_1^{(0)}, \dots,  \gamma_p^{(0)}\big)$. \\ 
				Proposed starting model is the full one with $\gamma_j^{(0)}=1$ for all $j=1,\dots,p$.} 
			\vspace{-4ex}
			\notation{$M_1$: current model in $t$ iteration; $M_2$: proposed model in $t$ iteration; $M_0$: reference model of PEP prior} 
			
			\Output{Random sample $g^{(t)}$ and $\dn{\gamma}^{(t)}$ for $t=1,\dots T$}

			\For{$t=1$ \KwTo $T$}{
				--- Set $M_1$ the model which corresponds to $\dn{\gamma}^{(t-1)}$\\ 
				\blocknotext(--- Generate random values of $g$:){Sample random values of $g^{(t)}$ from $f(g| \dn{y}, M_1)$ given by (17) in the main paper.}
				\vspace{-2em}
				\For{$j=1$ \KwTo $p$}{
					\blocknotext(BVS step for covariate $X_j$ conditional on $g$:){ 
						Implement BVS step of Algorithm \ref{algo0} with acceptance probability given by
						$$
						\alpha = \min \left\{ 1, \frac{f(\dn{y}|g^{(t)}, M_2) \pi_2\big(g^{(t)}\big) \pi(M_2)}
						{f(\dn{y}|g^{(t)}, M_1) \pi_1\big(g^{(t)}\big) \pi(M_1)} \right\} 
						$$				    		
						with the marginal likelihoods $f(\dn{y}|g, M)$, conditionally on $g$, given by (15) in the main paper,  
						$\pi_M(g)$ is the hyper-prior of $g$ under model $M$ and $\pi(M)$ is the prior  probability of model $M$; $M \in \{ M_1, M_2\}$. 
					}
					\vspace{-2em}
				}	
			}
			\blocknotext(END of the algorithm){}
			\vspace{-2em}
			\caption{Model search using $MC^3$ conditional on $g$}
			\label{algo2}
		}
	}
\end{algorithm}

\begin{algorithm} 
\adjustbox {scale=0.9}{ 	
	\colorbox{gray!25}{\parbox{0.90\textwidth}{
			\SetKwInOut{notation}{Notation}
			\SetKwInOut{Input}{Input}
			\SetKwInOut{Output}{Output}
			\SetKwBlock{blocknotext}{~}{~}
			\SetKwBlock{step}{Step}{~}
			\SetKwBlock{mcmcmcstep}{Description of an $MC^3$ Step}{~}
			\SetKwBlock{ini}{Initialization}{~}
			\SetKwBlock{schemeb}{SA version 1: Full SA (Scheme B)}{End-Full-SA}
			\SetKwBlock{schemec}{SA version 2: Partial SA (Scheme C)}{End-Partial-SA}
			\SetKwFor{For}{for}{~}{endfor}
			\SetKwFor{ForAll}{for~all}{~}{endfor}
			\SetKwFor{ForEach}{for~each}{~}{endfor}
			\footnotesize  
			\ini({\bf 1:} Use variable inclusion indicators $\dn{\gamma}$ as in Algorithm \ref{algo1}){~}	
			\vspace{-5ex}
			\ini({\bf 2:} Define the full parameter vector $\dn{\beta}$:){$\dn{\beta} =  (\beta_0, \beta_1, \dots, \beta_p)$.}	
			\vspace{-5ex}
			\ini({\bf 3:} Set initial values of $\dn{\beta}$, $\sigma$, $g$ and $\dn{\gamma}$){Set 
				$\dn{\beta}^{(0)}$, $\sigma^{(0)}$, $g^{(0)}$ and $\dn{\gamma}^{(0)}$ \\ 
				Proposed starting model is the full one with $\beta_j^{(0)}=0$, $\sigma^{(0)}=1$, $\gamma_j^{(0)}=1$ for all $j=1,\dots,p$ and proposed $g^{(0)}=n$.} 
						\vspace{-4ex}
			\notation{$M_1$: current model in $t$ iteration; $M_2$: proposed model in $t$ iteration; $M_0$: reference model of PEP prior; $\dn{\beta}$ parameter vector which contains the actual model parameters $\dn{\beta}_1$ and the rest of parameters; $\sigma^2$ denotes each model specific variance $\sigma_\ell^2$ for each model $M_\ell$} 
			
			\Output{Random sample $\dn{\beta}^{(t)}$, $\sigma^{(t)}$, $g^{(t)}$ and $\dn{\gamma}^{(t)}$ for $t=1,\dots T$}

			\For{$t=1$ \KwTo $T$}{
				--- Set $M_1$ the model which correspondings to $\dn{\gamma}^{(t-1)}$\\ 
				\blocknotext(--- Generate random values of $\dn{\beta}$:){
					--- Sample random values of $\beta_j^{(t)}$ included in $M_0$ from (13) in the main paper\\
					--- Sample random values of $\beta_j^{(t)}$ included in $M_1$ but not in $M_0$ from (12) in the main paper\\ 
					--- Sample random values of $\beta_j^{(t)}$ not included in $M_1$ from a ``pseudoprior'' 
					$f(\beta_j | \gamma_j=0)$. 
				}
				\vspace{-2em} 
				\blocknotext(--- Generate random values of $\sigma^2$:){Sample random values of ${\sigma^{2}}^{(t)}$ from $\pi_1(\sigma^2_1 | \dn{\beta}, g, \dn{y})$ given by (14) in the main paper.}
				\vspace{-2em} 
				\blocknotext(--- Generate random values of $g$:){Sample random values of $g^{(t)}$ from $\pi_1(g| \dn{y})$ given by (17) in the main paper.}
				\vspace{-2em}
				\For{$j=1$ \KwTo $p$}{
					\blocknotext(BVS step for covariate $X_j$ conditional on $g$:){ 
						Implement BVS step of Algorithm  \ref{algo0} with acceptance probability given by
						$$
						\alpha = \min \left\{ 1, 
						\frac{f(\dn{y}|\dn{\beta}^{(t)}, {\sigma^2}^{(t)}, M_2) 
							\pi_2\big(\dn{\beta}^{(t)}, {\sigma^2}^{(t)} \big| g^{(t)}\big) \pi_2\big(g^{(t)}\big)\pi\big(M_2\big)}
						{f(\dn{y}|\dn{\beta}^{(t)}, {\sigma^2}^{(t)}, M_1) 
							\pi_1\big(\dn{\beta}^{(t)}, {\sigma^2}^{(t)} \big| g^{(t)}\big) \pi_1\big(g^{(t)}\big)\pi\big(M_1\big)}  \right\} 
						$$				    		
						with 
						$f(\dn{y}|\dn{\beta}, {\sigma^2}, M) $ denoting the usual normal likelihood under model $M$ given the model parameters $\dn{\beta}$ and $\sigma^2$, 
						$\pi_M(\dn{\beta}, {\sigma^2} | g)$ is the prior of model parameters $\dn{\beta}$ and $\sigma^2$ for model $M$ given $g$,  
						$\pi_M(g)$ is the hyper-prior of $g$ under model $M$ 
						and $\pi(M)$ is the prior probability of model $M$; $M \in \{ M_1, M_2\}$. 
					}
					\vspace{-2em}
				}	
			}
			\blocknotext(END of the algorithm){}
			\vspace{-2em}
			\caption{Model search using a full MCMC algorithm -- Gibbs based variable selection sampler }
			\label{algo3}
		}
	}}
\end{algorithm}

\clearpage
\newpage 
\section{SDM Dataset: Further Results} 
\label{sdm_more}

\subsection{Posterior variable inclusion probabilities and posterior distributions of the model dimension}

\begin{figure}[htb!]
	\centering{}
	\includegraphics[scale=0.56]{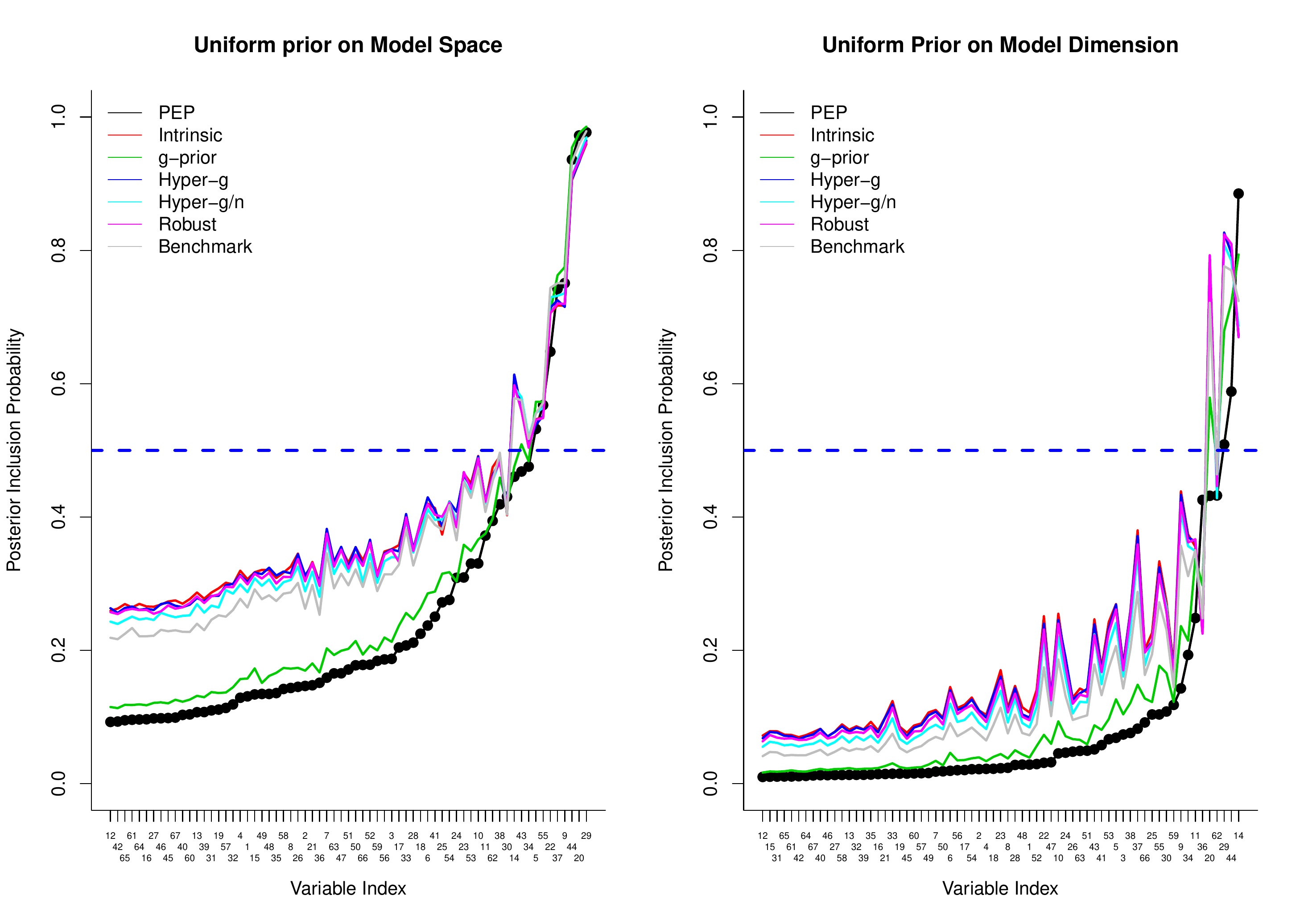}
	\caption{SDM Dataset: Posterior inclusion probabilities for all covariates (100K MCMC iterations).}
	\label{sdm_incprobs}
\end{figure}

In Figure \ref{sdm_incprobs} we present the posterior inclusion probabilities, based on the MCMC runs, for all covariates, under all competing methods, using the uniform prior on model space and the uniform prior on model size. The variables have been sorted, on the x-axis, according to their resulting posterior inclusion probabilities under the PEP prior; the horizontal dashed lines indicate posterior inclusion probabilities of 0.5. Under both prior setups for the model space, the inclusion probabilities for all covariates almost coincide under the robust, the intrinsic and the hyper-$g$ priors.
Under all competing approaches, the majority of the explanatory variables have posterior inclusion probabilities below 0.5. Focusing on those variables, under the PEP prior and the $g$-prior, the posterior inclusion probabilities are lower than the ones obtained using the rest of the methods. Those differences are more noticeable when using the uniform prior on model space. On the other hand, there are negligible differences, especially under the uniform prior on model space, between the reported posterior inclusion probabilities under all methods that greatly exceed the value 0.5. 
Finally there is a small number of variables with posterior inclusion probabilities below 0.5 under the PEP prior and the $g$-prior and above 0.5 under the rest of the methods. Those differences are more sharp when using the uniform prior on the model space.  
All the above findings imply that, regardless the prior used on the model space, the PEP prior, followed by the $g$-prior, are more parsimonious than their competitors.

\begin{figure}[htb!]
	\centering{}
	\includegraphics[scale=0.33]{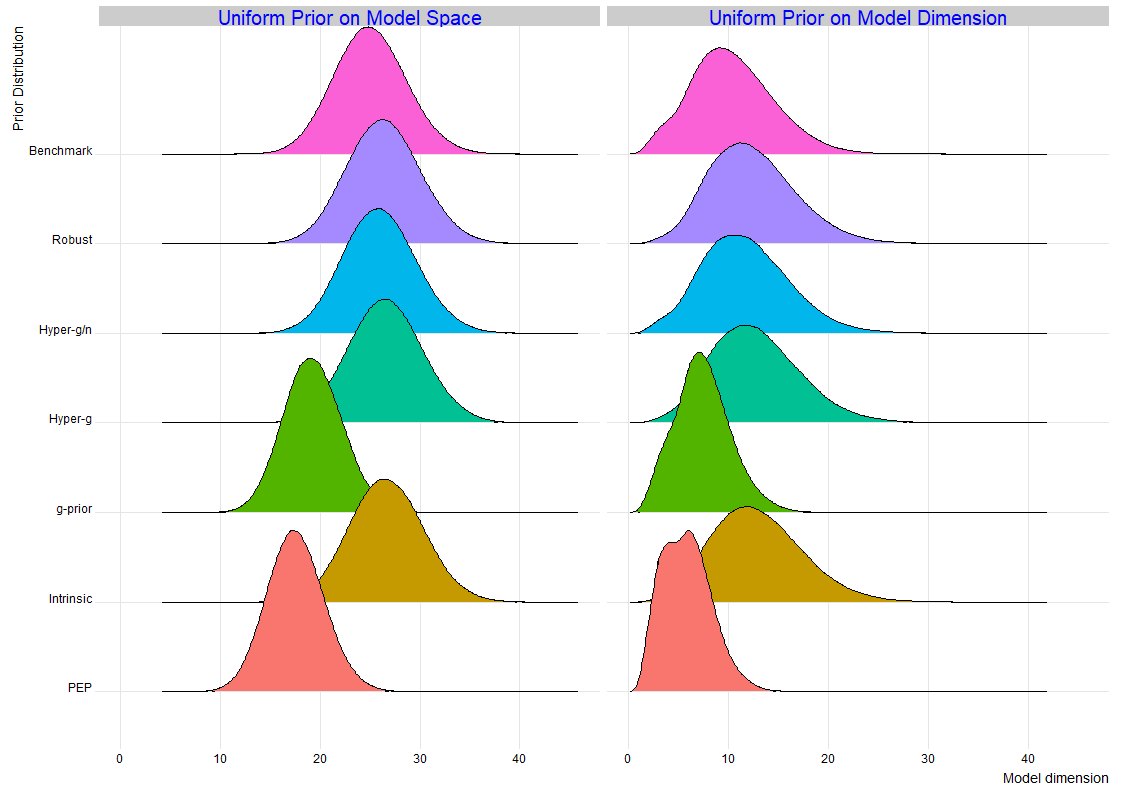}
	\vspace{-1.5em} 
	\caption{SDM Dataset: Posterior densities of model dimension for each method (100K MCMC iterations).}
	\label{sdm_boxplot_dim}
\end{figure}

In Figure \ref{sdm_boxplot_dim} we present the posterior densities of the model dimension for each competing method and each prior on the model space, based on the MCMC runs. Under both model space priors, the posterior densities of the model dimension almost coincide under the robust, the intrinsic, the hyper-$g$ and hyper-$g/n$ and the benchmark priors. All methods visited models with smaller dimensions on average under the uniform prior on model size. Furthermore, regardless the prior used on the model space, the PEP prior, followed by the $g$-prior, visited models with lower dimension on average and with posterior densities having a slightly smaller variance. Therefore, we come to the conclusion, for once more, that regardless the prior used on the model space, the PEP prior, followed by the $g$-prior, are more parsimonious than their competitors. 

\newpage
\subsection{BMA estimates of $R^2$ and RMSE}

Tables \ref{R2BMA}--\ref{RMSEBMA} present the BMA point estimates of  $R^2$ and RMSE, respectively, based on the posterior mean.

 \begin{table}[htb!]
	\caption{SDM Dataset: BMA estimates of $R^2$ obtained by the posterior mean of Gibbs based variable selection samplers}
	\label{R2BMA}
	\begin{center}
		\footnotesize
		\begin{tabular}{lccccccc} 
			\hline 
			&  PEP  &Intrinsic &$g$-prior& Hyper-$g$& Hyper-$g/n$  & Robust & Benchmark \\ 
			\hline                      
			Uniform on models    & 0.803 &    0.743 &  0.808  & 0.748    & 0.763        & 0.746  &   0.781 \\
			Uniform on dimension & 0.659  &   0.701 &  0.687  & 0.703    & 0.707        & 0.702  &  0.704 \\ 
			\hline 
		\end{tabular} 
		\normalsize 
	\end{center}
\end{table}

\begin{table}[htb!]
	\caption{SDM Dataset: BMA estimates of RMSE obtained by the posterior mean of Gibbs based variable selection samplers}
	\label{RMSEBMA}
	\begin{center}
		\footnotesize
		\begin{tabular}{lccccccc} 
			\hline 
			&  PEP  &Intrinsic &$g$-prior& Hyper-$g$& Hyper-$g/n$  & Robust & Benchmark \\ 
			\hline                      
			Uniform on models    & 1.23  & 1.34  & 1.21  & 1.33  & 1.30  & 1.34  & 1.27 \\ 
			Uniform on dimension & 1.57  & 1.45  & 1.51  & 1.45  & 1.44  & 1.45  & 1.46 \\
			\hline 
		\end{tabular} 
		\normalsize 
	\end{center}
\end{table}

\subsection{BMA posterior boxplots of log-predictive scores over 8-fold CV}

\begin{figure}[htb!]
	\centering{}
	\includegraphics[scale=0.4]{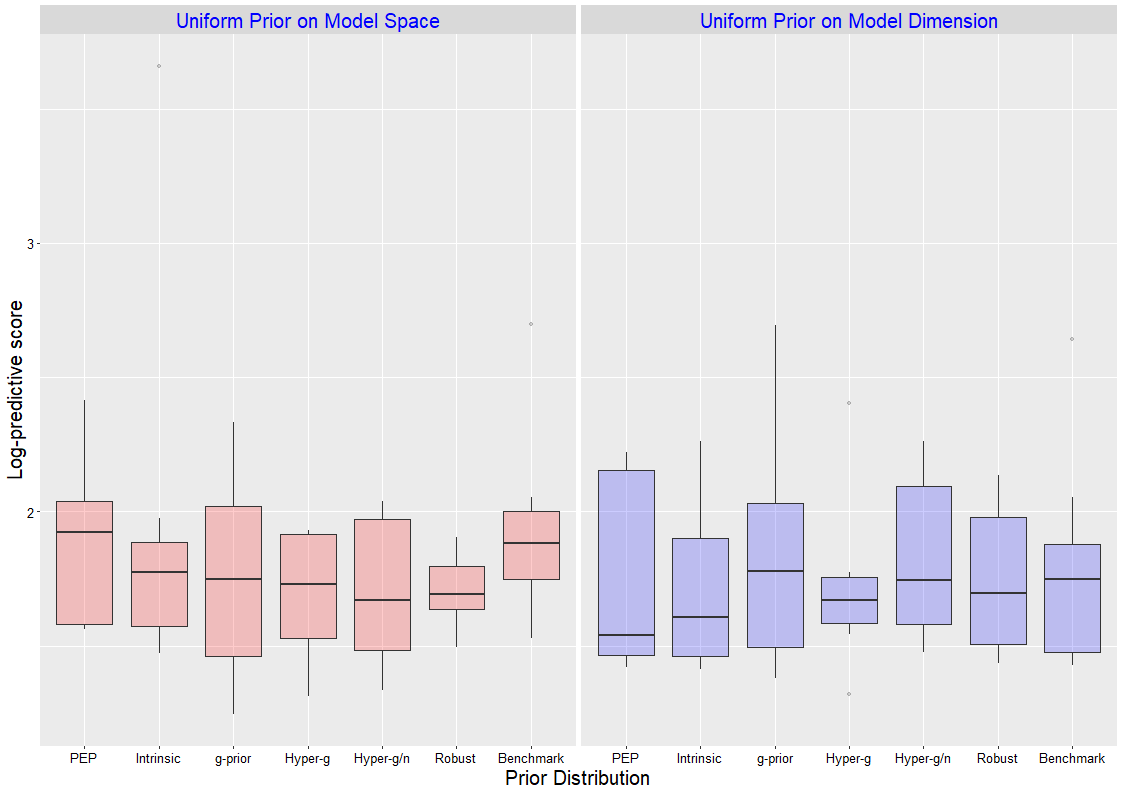}
	\vspace{-1.5em} 
	\caption{SDM Dataset: BMA posterior boxplots of log-predictive scores over 8-fold CV.}
	\label{sdm_boxplot_logpredscore}
\end{figure}

Figure \ref{sdm_boxplot_logpredscore} presents the BMA posterior boxplots of log-predictive scores over 8-fold CV, for the SDM dataset.

\bibliographystyle{agsm}

\bibliography{biblio3}